\documentclass[aps,nofootinbib,superscriptaddress, showpacs,preprintnumbers,  nofootinbibt,referee]{revtex4-2}
\usepackage{epsfig}
\usepackage{multirow}
\usepackage{eurosym}
\usepackage{dcolumn}
\usepackage{bm}
\usepackage{enumerate}
\usepackage{float}
\usepackage{epstopdf}
\usepackage{amsmath}
\usepackage{bm}
\usepackage{amsfonts}
\usepackage{amssymb}
\usepackage{graphicx}
\usepackage{alphalph,mathtools}
\usepackage{etoolbox}
\usepackage{color}
\usepackage{booktabs}
\usepackage{footnote}
\usepackage{makecell,tabularx}
\usepackage[bottom]{footmisc}

\def\be{\begin{equation}}
\def\ee{\end{equation}}
\def\bea{\begin{eqnarray}}
\def\eea{\end{eqnarray}}

\begin{document}

\title{Jacobi and Lyapunov stability analysis of circular geodesics around a spherically symmetric dilaton black hole}
\author{Cristina Blaga}
\email{cristina.blaga@ubbcluj.ro}
\affiliation{Faculty of Mathematics and Computer Sciences, Babes-Bolyai University, Cluj-Napoca, Romania,}
\author{Paul Blaga}
\email{aurel.blaga@ubbcluj.ro}
\affiliation{Department of Mathematics and Computer Sciences, Babes-Bolyai University, Cluj-Napoca, Romania, }
\author{Tiberiu Harko}
\email{tiberiu.harko@ubbcluj.ro}
\affiliation{Department of Physics, Babes-Bolyai University, Kogalniceanu Street,
	Cluj-Napoca, 400084, Romania,}

\begin{abstract}
We analyze the stability of the geodesic curves in the geometry of the Gibbons-Maeda-Garfinkle-Horowitz-Strominger black hole, describing the space time of a charged black hole in the low energy limit of the string theory.  The stability analysis is performed by using both the linear (Lyapunov) stability method, as well as the notion of Jacobi stability, based on the Kosambi-Cartan-Chern theory. Brief reviews of the two stability methods are also presented. After obtaining the geodesic equations in spherical symmetry, we reformulate them as a two-dimensional dynamic system.  The Jacobi stability analysis of the geodesic equations is performed by considering the important geometric invariants that can be used for the description of this system (the nonlinear and the Berwald connections), as well as the deviation curvature tensor, respectively. The characteristic values of the deviation curvature tensor are specifically calculated, as given by the second derivative of effective potential of the geodesic motion. The Lyapunov stability analysis leads to the same results. Hence, we can conclude that in the particular case of the geodesic motion on circular orbits in the Gibbons-Maeda-Garfinkle-Horowitz-Strominger, the Lyapunov and the Jacobi stability analysis gives equivalent results.
\end{abstract}

\date{\today }
\maketitle
\tableofcontents

\section{Introduction}

The analysis of the global stability of the solutions of the systems of strongly nonlinear, first or second order differential equations, describing the temporal evolution of complicated dynamical systems,  is usually performed, in the framework of a rigorous mathematical approach, with the help of the well known, and mathematically extensively investigated Lyapunov stability theory. In this commonly accepted approach to the problem of the stability of the solutions of differential equations, the basic mathematical quantities to be studied are the Lyapunov exponents. The Lyapunov exponents determine the exponential departures of the trajectories, obtained as solutions of the differential equations describing a given
dynamical system, as compared to a standard trajectory, taken as reference \citep{1,2}. But when
one attempts to obtain the Lyapunov exponents, one must understand that generally
it is a very complicated task to find them in an exact analytical form. For that reason, in
order to evaluate the basic Lyapunov exponents, and their properties, one should apply mostly complicated numerical methods.
Currently, a large number of powerful numerical methods have been established for their calculation.
Such numerical methods are extensively used for the mathematical, as well as physical description of the temporal evolution near the critical points of the differential equations describing dynamical systems \cite{3,12}.

The mathematical approach based on the Lyapunov linear stability analysis is well developed from a mathematical point of view. Moreover, it
provides an intuitive and clear comprehension of some of the stability properties of the systems of differential equations. Nevertheless, to obtain a more profound understanding of the time evolution and general properties of the natural and mathematical structures, alternative approaches for the investigation of the
stability problems must also be proposed, examined, and fully investigated. Hence, after a new
method for the study of the stability is developed, one could contrast exhaustively the results, predictions, and the possible shortcomings of the newly introduced method with the consequences and conclusions gained by adopting the
linear Lyapunov stability analysis of the considered system of differential equations.

One of the alternative approaches to the problem of stability that could provide valuable information for the investigation
of the stability of the systems of ordinary second order differential equations is given by what one could
designate as the geometro-dynamical approach. Historically, one of the first illustrations of such a
different mathematical investigation of the stability problem of the systems of differential equations is the Kosambi-Cartan-Chern (KCC) theory.
In its early formulation the KCC theory was developed in the major investigations of
Kosambi \cite{Ko33}, Cartan \cite{Ca33} and Chern \cite{Ch39}, respectively, which led to a first rigorous mathematical investigation of the geometric properties of dynamical systems.
From a strictly  mathematical perspective, the KCC theory is strongly influenced by the geometry
of the Finsler spaces, which largely represents, and significantly contributed, to its theoretical justification. The KCC
theory is built up by using the basic conjecture according to which there is a mathematical analogy
between autonomous or non-autonomous dynamical system, formulated in terms of second order differential equations, and
the equations of the geodesics in a Finsler space. The Finsler type geodesic equations can always be associated to a given dynamical system, described by second order, usually strongly nonlinear systems of differential equations  (for an in depth presentation of the KCC theory, and of its applications, see \cite{rev}).

The KCC theory represents essentially a geometric way of thinking about
the variational differential equations describing the divergence/convergence of a bunch
of trajectories of a dynamical system, or of a system of differential equations, as compared to the neighbouring ones \cite%
{An00}. The KCC theory proposes a geometrical type characterization of the systems of second order differential equations, considered as geodesic curves. Within the framework of this description one can construct for each system of differential equations two connections, which are essentially geometric quantities. The first connection is the
non-linear connection $N_j^i$, while a Berwald type connection $G^i_{jk}$ is also considered  for the description of the dynamical system.
With the use of these two connections, which can be generally defined, five important geometrical invariants can be constructed rigorously. Of
these five geometrical invariants, the most relevant, from a mathematical and physical point of view,  is the second invariant $P^i_j$, which is named the
deviation curvature tensor. From the general perspective  of the
scientific, engineering and mathematical applications, its importance is given by its main property of determining
the so-called Jacobi stability of the considered system of second order nonlinear differential equation  \cite{rev,
An00, Sa05,Sa05a, An93}. Various engineering, physical, chemical, biochemical, or medical systems
have been thoroughly investigated with the help of the KCC stability theory \cite{YaNa07, Har1, Har2, T0, T1, X1, X2, X3, X4, Ha3, Ha4, Ha4a, Ha5, Ha6, BBH}.

The KCC theory has also found applications in the study of the gravitational
phenomena. Thus, in \cite{Har1}, the static, spherically symmetric structure equations of
the static vacuum in the brane world models were analyzed, from the point of view of their stability, by applying both the Jacobi stability analysis, and the linear
(Lyapunov) stability analysis. It was
shown that the trajectories that are unstable on the static brane with spherical symmetry behave chaotically, which implies
that after a bunch of particles travel a restricted range of a radial distance, it would not be possible to differentiate
the trajectories that were extremely close to each other at an initial time, and at an initial point. Thus,
the KCC theory together with the Jacobi stability analysis represents a very powerful method for giving some important constraints on the
physical properties of the vacuum on the four-dimensional brane.

The stability of the radial
solutions of the Lane-Emden semilinear elliptic equation $\Delta u+u^n=0$,
with the initial conditions $u(0)=1$ and $u^{\prime }(0)=0$, respectively,  were studied on the positive real line
in \cite{Har2}, by using the Lyapunov standard linear stability analysis, the Jacobi stability
approach, and the Lyapunov function method, respectively. By using the KCC
theory one can obtain for the stability of a polytropic star the criterion $%
E_i/E_g<n\rho(r)/\bar{\rho}$, where $E_i$ is the internal energy, $E_g$ is
the gravitational energy, $\rho(r)$ is the density of the stellar matter, and $\bar{\rho}%
(r) $ is the mean mass density, respectively. An investigation of the stability properties of the inflationary cosmological  models in the presence of scalar fields, by using the Jacobi stability analysis, was performed by considering a "second geometrization" of the models, and interpreting them as paths of a semispray, in \cite{Ha4}. The KCC stability properties of the cosmological models in the presence of scalar fields with exponential and Higgs type potentials were considered in detail. The KCC theory was used to investigate the Jacobi type stability/instability of the string equations in \cite{Ha5}. Moreover, by using this approach,  precise bounds on the geometrical and physical parameters that guarantee dynamical stability of the windings were determined. It was found that for the same initial conditions, and in higher dimensions, the topology and the curvature of the internal space have significant influences on the microscopic behavior of the string. On the other hand, it turns out, surprisingly, that the macroscopic behavior of the string is not sensitive to the details of the physical motion in the compact space. The stability of the circular restricted three
body problem, which considers the motion of a particle with a very small mass due to
the gravitational attraction of two massive stellar type objects moving on circular orbits
about their common center of mass, was analyzed in \cite{BBH}, by using the
KCC) theory. It was found that from the geometric perspective of the KCC theory, the
five Lagrangian equilibrium points of the restricted three body problem are
all unstable.

Black holes are important observational astrophysical objects, as well as
fundamental testing grounds for the theories of gravitation. In particular,
the motion of particles around the central black hole can give essential
information on the physical processes taking place in the cosmic
environments. There are many known black hole solutions, obtained in the
framework of the different gravitational theories. Exact black hole type
solutions of string theory play an important role for the confrontation of
the predictions of the theory with observations. Static, spherically
symmetric charged black hole solutions in the low energy limit of string
theory were found in \cite{GM88} and \cite{GHS91}, respectively. These
solutions are characterized by three independent parameters: their gravitational mass, electric charge, and the asymptotic value
of the scalar dilaton field, respectively, whose presence has important physical consequences. A particular class of solutions, the
extremely charged "black holes", represent, from a geometric point of view,  geodesically complete
spacetimes, without event horizons and singularities.

It is the goal of the present paper to study the stability properties
of the geodesic trajectories in the charged dilatonic solution of the low
energy limit of string theory, as obtained in \cite{GM88} and \cite{GHS91},
respectively. After obtaining the geodesic equations of motion, and the
expression of the effective potential, the stabiligty of the trajectories is
analyzed by using both the linear Lyapunov, and the KCC theory based Jacobi stability approaches. It
turns out that in the case of the string theory inspired dilatonic black
hole solution the predictions of both stability methods coincide.

The present paper is organized as follows. We briefly review the Lyapunov
and the Jacobi stability approaches in Section~\ref{sect1}. The charged
black hole solution of the dilatonic low energy string theory is written
down in Section~\ref{sect2}, where the equations of the geodesic lines are
also obtained. The stability of the trajectories of the particles moving
around the black hole are investigated, by using both the Lyapunov and the
Jacobi approaches, in Section~\ref{sect3}. We discuss and conclude
our results in Section~\ref{sect4}.

\section{Brief review of Lyapunov stability, and of the KCC theory}
\label{sect1}

In the present Section we quickly introduce, in a concise but rigorous way, the fundamental ideas, the basic
concepts,  and the results of the linear Lyapunov stability theory, and of the KCC theory,
respectively. For in depth discussions of the mathematical aspects of the linear (Lyapunov) stability theory, and of its applications in astrophysics and cosmology see \cite{cosm1,cosm2,cosm3,cosm4}.

Furthermore,we present the notations of the geometrical and physical quantities used in the present investigation, and we introduce the basic definitions
of the relevant geometric objects met in KCC theory (for a detailed
presentation we refer the reader to \cite{rev} and \cite{An00}).

\subsection{Linear stability of the systems of Ordinary Differential Equations}

We start by concisely presenting, mostly using the approach introduced in \cite{M}, some key consequences of the
Lyapunov stability analysis of arbitrary dynamical systems, expressed by general systems of first order
Ordinary Differential Equations (ODEs). In beginning our presentation we would like to mention that
the stability of a system of first ordinary differential equations is determined, in general, by the roots of its
characteristic polynomial. To prove this property, we consider the system of autonomous first order ordinary differential equations \cite{M},
\begin{eqnarray}  \label{1}
\frac{dx^{1}}{dt} &=&f^{1}\left( x^{1},x^{2},...,x^{n}\right) ,  \notag \\
\frac{dx^{2}}{dt} &=&f^{2}\left( x^{1},x^{2},...,x^{n}\right) ,  \notag \\
&&..............., \\
\frac{dx^{n-1}}{dt} &=&f^{n-1}\left( x^{1},x^{2},...,x^{n}\right) ,  \notag
\\
\frac{dx^{n}}{dt} &=&f^{n}\left( x^{1},x^{2},...,x^{n}\right) ,  \notag
\end{eqnarray}
where $f_1, f_2, \dots , f_n$ are, by definition, $n$ smooth functions, possessing derivatives of all orders in their domain of definition. Now let us linearize the
 system (\ref{1}) about one of its steady states (fixed point, or equilibrium point) $\left(
x_{0}^{1},x_{0}^{2},...,x_{0}^{n}\right) $, by associating to the arbitrary system (\ref{1}) the linear system
\begin{equation}
\begin{pmatrix}
\frac{dx^{1}}{dt} \\
\frac{dx^{2}}{dt} \\
... \\
\frac{dx^{n}}{dt}%
\end{pmatrix}%
=A%
\begin{pmatrix}
x^{1} \\
x^{2} \\
... \\
x^{n}%
\end{pmatrix}%
,  \label{2}
\end{equation}%
where the Jacobian matrix $A$ of the
system (\ref{1}), defined as $A=\left. J\left( f^{1},f^{2},...,f^{n}\right) \right\vert _{\left(
x_{0}^{1},x_{0}^{2},...,x_{0}^{n}\right) }$, is  evaluated at the steady state,
\begin{equation*}
A=\left.
\begin{pmatrix}
\frac{\partial f^{1}}{\partial x^{1}} & \frac{\partial f^{1}}{\partial x^{2}}
& ... & \frac{\partial f1}{\partial x^{n}} \\
... & ... & ... & ... \\
\frac{\partial f^{n}}{\partial x^{1}} & \frac{\partial f^{n}}{\partial x^{2}}
& ... & \frac{\partial f^{n}}{\partial x^{n}}%
\end{pmatrix}%
\right\vert _{\left( x_{0}^{1},x_{0}^{2},...,x_{0}^{n}\right) }.
\end{equation*}

The solutions of Eq.~(\ref{2}) can be obtained as \cite{M}
\begin{equation}
\begin{pmatrix}
x^{1} \\
x^{2} \\
... \\
x^{n}%
\end{pmatrix}%
=%
\begin{pmatrix}
C^{1}e^{\lambda _{1}t} \\
C^{2}e^{\lambda _{2}t} \\
... \\
C^{n}e^{\lambda _{n}t}%
\end{pmatrix}%
,  \label{3}
\end{equation}%
where by $\left( C^{1},C^{2},...,C^{n}\right) $ we have denoted the $n$ components of an arbitrary  vector quantity having constant components, while $%
\left( \lambda _{1},\lambda _{2},...,\lambda _{n}\right) $ denote the proper values
(eigenvalues) of the matrix $A$, which are determined as the algebraic roots of the characteristic
polynomial $p(\lambda)$, which are defined according to the relation,
\begin{equation}
\det \left( A-\lambda I_{n}\right) =0,
\end{equation}%
where by $I_{n}$ we have denoted the identity matrix, defined in the standard way.

\textbf{Definition} \cite{M}. \textit{Let's assume that a solution of the system of the ordinary differential equations (\ref{1}) is known. The solution is
called stable if and only if all the roots $\left(\lambda _{1}, \lambda _{2},
..., \lambda _{n}\right)$, of the characteristic polynomial $p(\lambda) $ are located on
the left hand side of the complex plane, that is, for all roots $\lambda $ the condition $\mathrm{Re} \lambda <0$ is satisfied.}

Assuming that the condition $\mathrm{Re}\;\lambda <0$ is satisfied, then it follows that as  $t\rightarrow \infty $, $x^{i}(t)=e^{\lambda _{i}t}$, $i=1,2,...,n$, tend
exponentially to zero  for all $i$. Hence, the point  $\left(
x^{1},x^{2},...,x^{n}\right) =\left( 0,0,...,0\right) $ is stable with respect to small
(linear) perturbations of the system of differential equations.

In the case of $n$-dimensional system of ordinary differential equations, the characteristic polynomial is given by
\begin{equation}
\mathbf{p}(\lambda )=\lambda ^{n}+a_{1}\lambda ^{n-1}+...+a_{n}=0,
\label{pol}
\end{equation}%
where the coefficients $a_{i}$, are all real numbers, $a_i\in R^n$, $i=1,2,...,n$. Moreover, without any
loss of generality, it is possible to consistently consider $a_{n}\neq 0$, since otherwise we would
obtain $\lambda =0$, and thus we would have a characteristic polynomial of order $n-1$,
having the coefficient of the zeroth order non-vanishing.

The important, and fundamental necessary and sufficient conditions for the polynomial $\mathbf{p}\left( \lambda
\right) $ to have all algebraic solutions satisfying the condition $\mathrm{Re}\lambda <0$ may be generally formulated as \cite{M}
\begin{equation*}
a_{n}>0, D_{1}=a_{1}>0, D_{2}=%
\begin{vmatrix}
a_{1} & a_{3} \\
1 & a_{2}%
\end{vmatrix}%
>0,
\end{equation*}

\begin{equation}
D_{3}=%
\begin{vmatrix}
a_{1} & a_{3} & a_{5} \\
1 & a_{2} & a_{4} \\
0 & a_{1} & a_{3}%
\end{vmatrix}%
>0,..., D_{k}=%
\begin{vmatrix}
a_{1} & a_{3} & ... & ... \\
1 & a_{2} & a_{4} & ... \\
0 & a_{1} & a_{3} & ... \\
0 & 1 & a_{2} & ... \\
... & ... & ... & ... \\
0 & 0 & ... & a_{k}%
\end{vmatrix}%
>0,
\end{equation}%
for all $k=1,2,...,n$.

An alternative, and important method for the study of the linear, Lyapunov stability question, is to investigate the algebraic relations
involving the non-zero solutions of the characteristic polynomial $\mathbf{p}\left( \lambda \right) $. Thus, we obtain,
\begin{equation}
s:=\sum_{i=1}^{n}\lambda _{i}=-a_{1},
\end{equation}
\begin{equation}
\mu _{1}:=\sum_{i,j,i\neq j}^{n}\lambda _{i}\lambda _{j}=a_{2},
\end{equation}
\begin{equation*}
...
\end{equation*}
\begin{equation}
p:=\lambda _{1}...\lambda _{n}=\left( -1\right) ^{n}a_{n}.
\end{equation}

From the values of these coefficients, and by taking into account their algebraic properties,  we can determine
essential and significant information on the stability properties of the system of ordinary differential equations (\ref{1}). in the following, for the sake of completeness of our discussion, we also present the

\textbf{Remark} (Descartes' Rule of Signs) \cite{M}. Let us consider the characteristic polynomial (\ref{pol}) of the system of differential equation (\ref{1}), with the coefficient $a_n$ satisfying the condition $a_{n}>0$. Let us denote by $m$ the number of changes in sign in the
sequence of the coefficients $\left\{ a_{n},a_{n-1},...,a_{0}\right\} $,
ignoring any coefficients in the sequence that are zero. Then, the polynomial $\mathbf{p}(\lambda )$ has at most $m$ roots
 that are real and positive. In addition,  $m$%
, $m-2$, $m-4$, ..., real positive roots of the polynomial do exist \cite{M}.

By taking $\omega :=-\lambda $, Descartes' Rule of Signs gives essential clues
about the potential existence of real negative roots of the characteristic polynomial, an information that is crucial for the study of the
stability of the systems of strongly nonlinear differential equations.

If the proper values of the Jacobian $A$ associated to a system of differential equations, evaluated at the equilibrium point $%
x_0:=\left( x_{0}^{1},x_{0}^{2},...,x_{0}^{n}\right)$ are known, with Eq.~(%
\ref{3}), we get the behavior of solution near $x_0$. For example, if we
consider a two dimensional autonomous differential system we obtain the
following classification of the equilibrium points: if the eigenvalues of $A$
have negative real parts, then in the phase plane, all solutions are converging towards
the steady state (equilibrium point) $x_0$. The point $x_0$ is named hyperbolic
sink (stable point). Moreover, if the real parts of the proper values (eigenvalues) of $A$ are greater
then zero, then all integral curves diverge from the equilibrium point and $%
x_0$ is named hyperbolic source (unstable point). If one eigenvalue is
positive, and the other one is negative, the fixed point $x_0$ is a saddle point
(unstable). If the proper values (eigenvalues) of $A$ are complex conjugate pairs, and $%
\mathrm{Re}\;\lambda \neq 0$, then the equilibrium point is a spiral point
(stable if $\mathrm{Re}\;\lambda <0 $, and unstable otherwise). If the proper values
(eigenvalues) of $A$ are purely imaginary values ($\mathrm{Re}\;\lambda =0$),
the fixed point is named a center.

If the eigenvalues of the linearized system~(\ref{2}) evaluated in $x_0$
have nonzero real parts, then the equilibrium point is said to be
hyperbolic. Otherwise is called nonhyperbolic. The relation between the
linear stability of the system~(\ref{1}) and its linearization~(\ref{2}) at
equilibrium points is given by the following Theorem.

\textbf{Theorem} (Hartman-Grobman) \cite{X3} \textit{Let us consider a system of ordinary differential equations $\dot{x}%
=f(x)$, $x \in R^n$, with the vector field $f$ being $C^1$. Let's assume that $\bar{x}$ is a
\emph{hyperbolic} fixed point of the considered system of differential equations. Then, a neighborhood of the point
$\bar{x}$, on which the flow is \emph{topologically equivalent} to the flow
of the linearization of the system of differential equations at $\bar{x}$ does always exist.}

For the sake of completeness we mention that the linear stability of the system~(\ref{1}) near a nonhyperbolic point
could be investigated with the aid of the Lyapunov function introduced through
the following Theorem and definition.

\textbf{Theorem} (Lyapunov stability theorem) \cite{X3} Let us consider
that a vector field $\dot{x}=f(x)$, $x \in R^n$ is given. Let $\bar{x}$ denote an
equilibrium point of the vector field $\dot{x}$. Moreover, let $\mathbf{\Sigma}:U \rightarrow R$
be a $C^1$ function, defined on some neighborhood $U$ of $\bar{x}$, and having the properties

i) $\Sigma \left(\bar{x}\right)=0$ and $\Sigma (x)>0$, if $x\neq \bar{x}$,

ii) $\dot{\Sigma }(x)\leq 0$ in $U-\bar{x}$.

Then, if the conditions i) and ii) are satisfied, the point $\bar{x}$ is stable. Furthermore, if

iii) $\dot{\Sigma}(x)<0$ in $U-\bar{x}$,
the point $\bar{x}$ is asymptotically stable.

The function $\mathbf{\Sigma}(x)$ from the above Theorem is called the \emph{Lyapunov
function}. It has the property that near the equilibrium point $x_0$, the
integral curves $f(x)$ are tangent to the surface levels of $\mathbf{\Sigma}$, or
they cross the surface level oriented towards their interior. In other words, $%
\nabla \mathbf{\Sigma}(x_0) \cdot f(x_0) \leq 0$.

\subsection{KCC stability theory}

In the following we introduce the basic ideas, and geometric concepts, of the KCC theory. Our presentation mostly follow the similar expositions of the theory in \cite{rev} and \cite{Ha5}, respectively.

\subsubsection{Geometrization of arbitrary dynamical systems}

We introduce first a set of dynamical variables $x^i$, $i=1,2,...,n$, assumed to be defined on a
real, smooth $n$-dimensional manifold $\mathcal{M}$. We denote in the following by $T\mathcal{M}$ the tangent bundle of $%
\mathcal{M}$. Usually, $\mathcal{M}$ is considered
as $R^n$, $\mathcal{M}=R^n$, and therefore $T\mathcal{M}=TR^n=R^n$ \cite{Ha4a}.

Let's consider now a particular subset $\Omega $ of the  $(2n+1)$ dimensional Euclidian
space defined as $R^{n}\times R^{n}\times R^{1}$. On $\Omega $ we assume the existence of a $2n+1$
dimensional coordinate system denoted $\left(x^i,y^i,t\right)$, $i=1,2,...,n$, where we have also introduced the notations
$\left( x^{i}\right) =\left( x^{1},x^{2},...,x^{n}\right) $, and $\left(
y^{i}\right) =\left( y^{1},y^{2},...,y^{n}\right) $, respectively. By $t$ we denote the ordinary
time coordinate. We define the coordinates $y^i$ as
\begin{equation}
y^{i}=\left( \frac{dx^{1}}{dt},\frac{dx^{2}}{dt},...,\frac{dx^{n}}{dt}%
\right) .
\end{equation}

A fundamental supposition in the KCC theory refers to the coordinate $t$, which can be interpreted physically as the time variable,  and which is assumed to be an
absolute invariant, which does not change in the coordinate transformations. Thus, by taking into account this assumption, on the base manifold $\mathcal{M}$  the
only allowed transformations of the coordinates are of the general form \cite{Ha4a},
\begin{equation}
\tilde{t}=t,\;\tilde{x}^{i}=\tilde{x}^{i}\left( x^{1},x^{2},...,x^{n}\right),\;
i=1 ,2,...,n .  \label{ct}
\end{equation}

In various mathematical applications of scientific importance, and interest, the equations of motion describing the evolution of natural or engineering systems
 are obtainable from a Lagrangian function $L$, which describes the state of the system, and is an application $L:T\mathcal{M}%
\rightarrow R$. The dynamical evolution of the system can be obtained with the help of the Euler-Lagrange equations, given by \cite{Ha4a}
\begin{equation}
\frac{d}{dt}\frac{\partial L}{\partial y^{i}}-\frac{\partial L}{\partial
x^{i}}=F_{i},i=1,2,...,n.  \label{EL}
\end{equation}%
For the specific case of mechanical systems, the quantities $F_{i}$, $i=1,2,...,n$, give the components of
the external force $\vec{F}$, which cannot be derived from a potential. By assuming that the Lagrangian $L$ is regular, by using some simple
calculations, we obtain the fundamental results that the Euler-Lagrange equations Eq.~(\ref{EL})
are equivalent mathematically to a complicated system of second-order ordinary, strongly nonlinear
differential equations, given by \cite{rev, Ha4a}
\begin{equation}
\frac{d^{2}x^{i}}{dt^{2}}+2G^{i}\left( x^{j},y^{j},t\right) =0,i=
1,2,...,n.  \label{EM}
\end{equation}%
As for the functions $G^i$, we assume that  each function $G^{i}\left( x^{j},y^{j},t\right) $ is $C^{\infty }$ in
a neighborhood of some initial conditions $\left( x _{0},y _{0},t_{0}\right)
$, defined in $\Omega $.


The key conjecture of the KCC theory is the following. Let's assume that an arbitrary system of strongly nonlinear
second-order ordinary differential equations of the form (\ref{EM}%
) is defined in a general form. Even if the Lagrangian function for the system is not known \textit{a priori}, we still
can investigate the evolution and the behavior of its trajectories by using techniques suggested by
the differential geometry of the Finsler spaces. This analysis can be carried
out due to the existence of a close similarity between the paths of the
 Euler-Lagrange system, and the geodesics in a Finsler geometry.

\subsubsection{The non-linear and Berwald connections, and the KCC invariants associated to a dynamical system}

To investigate from a geometrical perspective the mathematical properties of the dynamical system
described by the system of differential equations Eqs.~(\ref{EM}),  we first introduce
the nonlinear connection $N$, defined on the base manifold $\mathcal{M}$, and having the coefficients $N_{j}^{i}$ given by
 \cite{MHSS}
\begin{equation}  \label{ncon}
N_{j}^{i}=\frac{\partial G^{i}}{\partial y^{j}}.
\end{equation}

From a general geometric perspective, the nonlinear connection $N_{j}^{i}$ can be
characterized with the help of a dynamical covariant derivative $\nabla ^N$ by using the following
procedure. Let's assume that two vector fields $v$, and $w$, respectively, are given, with both vector fields defined over a
manifold $\mathcal{M}$. Next,  we define the covariant derivative $\nabla ^N$ of
$w$ as \cite{Punzi}
\begin{equation}  \label{con}
\nabla _v^Nw=\left[v^l\frac{\partial }{\partial x^l}w^i+N^i_j(x,y)w^j\right]%
\frac{\partial }{\partial x^i}.
\end{equation}

For $N_i^j(x,y)=\Gamma _{il}^j(x)v^l$, from Eq.~(\ref{con}) we directly
reobtain the definition of the standard covariant derivative for the particular case of
the Levi-Civita linear connection, as introduced usually in the Riemannian
geometry \cite{Punzi}.

Now we consider the open subset $\Omega \subseteq R^{n}\times R^{n}\times
R^{1}$ on which the system of differential equation (\ref{EM}) is defined, together with the coordinate transformations, defined by Eqs.~(\ref{ct}), and assumed to be non-singular. On this
mathematical structure we can define the KCC-covariant derivative of an arbitrary
vector field $\xi ^{i}(x)$ by means of the definition \cite{An93,An00,Sa05,Sa05a},
\begin{equation}  \label{KCC}
\frac{D\xi ^{i}}{dt}=\frac{d\xi ^{i}}{dt}+N_{j}^{i}\xi ^{j}.
\end{equation}

By taking $\xi ^{i}=y^{i}$, we obtain
\begin{equation}
\frac{Dy^{i}}{dt}=N_{j}^{i}y^{j}-2G^{i}=-\epsilon ^{i}.
\end{equation}
The contravariant vector field $\epsilon ^{i}$ is defined on the subset $%
\Omega $ of the Euclidian space, as one can see immediately from the above equation. It repreents the first KCC invariant. From a
general physical perspective, and within the mathematical formalism of the
classical Newtonian mechanics, $\epsilon ^{i}$, the first KCC invariant,  could be
understood as an external force, not derivable from a potential, and acting on the dynamical system.

As a next step in our discussion of the KCC theory, we consider the infinitesimal variations of the trajectories $x^{i}(t)$ of the dynamical
system (\ref{EM}) into neighbouring ones, with the variations defined according to the prescriptions,
\begin{equation}  \label{var}
\tilde{x}^{i}\left( t\right) =x^{i}(t)+\eta \xi ^{i}(t), \tilde{y}^{i}\left(
t\right) =y^{i}(t)+\eta \frac{d\xi ^{i}(t)}{dt},
\end{equation}
where by $\left| \eta \right|<<1 $ we have denoted a small infinitesimal quantity, while $\xi ^{i}(t)$ denote
the components of an arbitrary contravariant vector field $\xi$. The vector field $\xi$ is defined along the
trajectory $x^{i}(t)$ of the system of the differential equations under consideration. After the substitution of Eqs.~(\ref%
{var}) into Eqs.~(\ref{EM}), and by considering the limit $\eta \rightarrow 0$, we
arrive to the deviation, or Jacobi, equations, representing the
central mathematical result of the KCC theory, and which are given by \cite%
{An93,An00,Sa05,Sa05a}
\begin{equation}  \label{def}
\frac{d^{2}\xi ^{i}}{dt^{2}}+2N_{j}^{i}\frac{d\xi ^{j}}{dt}+2\frac{\partial
G^{i}}{\partial x^{j}}\xi ^{j}=0.
\end{equation}

With the help of the KCC-covariant derivative, as introduced in Eq.~(\ref{KCC}), we can
reformulate Eq.~(\ref{def}) in a fully covariant form as
\begin{equation}
\frac{D^{2}\xi ^{i}}{dt^{2}}=P_{j}^{i}\xi ^{j},  \label{JE}
\end{equation}
where we have denoted
\begin{equation}  \label{Pij}
P_{j}^{i}=-2\frac{\partial G^{i}}{\partial x^{j}}-2G^{l}G_{jl}^{i}+ y^{l}%
\frac{\partial N_{j}^{i}}{\partial x^{l}}+N_{l}^{i}N_{j}^{l}+\frac{\partial
N_{j}^{i}}{\partial t}.
\end{equation}
In Eq.~(\ref{Pij}) we have defined the important tensor $G_{jl}^{i}$, given
by \cite{rev, An00, An93,MHSS,Sa05,Sa05a}
\begin{equation}
G_{jl}^{i}\equiv \frac{\partial N_{j}^{i}}{\partial y^{l}},
\end{equation}
and which in the KCC theory is named the Berwald connection.

The tensor $P_{j}^{i}$ is called the second KCC-invariant. It is the fundamental quantity in the KCC theory, and in the Jacobi stability investigations. Alternatively, $P_{j}^{i}$ is also called  the deviation curvature tensor, by indicating its essential geometric nature. Furthermore, we will call
in the followings Eq.~(\ref{JE}) as the Jacobi equation. This equation exists in both Riemann
or Finsler geometries.  If one assumes that the system of equations (\ref{EM}) corresponds to the
geodesic motion of a physical system, then Eq.~(\ref{JE}) gives the so-called Jacobi field equation, which
can always be introduced in the considered geometry.

 The trace $P=P_i^i$ of the curvature deviation tensor, constructed from $P^i_j$, is a scalar invariant. It can be calculated from the relation
\begin{equation}
P=P_{i}^{i}=-2\frac{\partial G^{i}}{\partial x^{i}}-2G^{l}G_{il}^{i}+ y^{l}%
\frac{\partial N_{i}^{i}}{\partial x^{l}}+N_{l}^{i}N_{i}^{l}+\frac{\partial
N_{i}^{i}}{\partial t}.
\end{equation}

Other important invariants can also be constructed in the KCC theory. The most commonly used invariants are the third, fourth and fifth
invariants associated to the given dynamical system, whose evolution and behavior is represented by the second order system of nonlinear equations (\ref{EM}). These
invariants are introduced according to the definitions \cite{An00}
\begin{equation}  \label{31}
P_{jk}^{i}\equiv \frac{1}{3}\left( \frac{\partial P_{j}^{i}}{\partial y^{k}}-%
\frac{\partial P_{k}^{i}}{\partial y^{j}}\right) ,P_{jkl}^{i}\equiv \frac{%
\partial P_{jk}^{i}}{\partial y^{l}},D_{jkl}^{i}\equiv \frac{\partial
G_{jk}^{i}}{\partial y^{l}}.
\end{equation}

From a geometrical perspective, $P_{jk}^{i}$, the third KCC invariant, can be described as a torsion tensor. $P_{jkl}^{i}$  the fourth KCC invariant, represents the equivalent of the Riemann-Christoffel curvature
tensor, while $D_{jkl}^{i}$, the fifth KCC invariant, is called the Douglas tensor \cite{rev, An00}. Let's point out now that in a Berwald geometry these three tensors can always be defined.
It is also important to mention that in the KCC theory, the five invariants defined above are the fundamental mathematical
quantities that describe the geometrical properties, and interpretation, of a dynamical system whose time evolution and behavior are represented by an
arbitrary system of second-order strongly nonlinear differential equations.

\subsubsection{Jacobi stability of dynamical systems}

In a large number of scientific investigations, the analysis of the stability of
biological, chemical, biochemical, engineering, medical  or physical systems,  as well as the study of the trajectories of a system of differential equations, as given, for example, by Eqs.~(\ref%
{EM}), in the vicinity of a given point $x^{i}\left( t_{0}\right) $, is of
fundamental significance for the understanding of its properties. Moreover, the study of the stability can give important information on
the temporal evolution of a general dynamical system,

For simplicity, in the following, we adopt as the origin of the time variable $t$ the value $t_{0}=0$. Moreover, we define $\left\langle
.,.\right\rangle $ as representing the canonical inner product of $R^{n}$.
We also introduce the null vector $O$ defined in $R^n$, $O\in R^{n}$. Next,
we assume that the trajectories $x^{i}=x^{i}(t)$ of the system of differential equations (%
\ref{EM}) represent smooth curves in the Euclidean space $R^{n}$, endowed with the canonical inner product $\left\langle
.,.\right\rangle $. Moreover, to completely characterize the deviation vector $\xi $, we assume, as a general property, that
it satisfies the set of two essential initial conditions $\xi \left( 0\right) =O$ and $\dot{\xi%
}\left( 0\right) =W\neq O$, respectively \cite{rev, An00, Sa05,Sa05a}.

To describe the dispersing/focusing trend of the trajectories of a dynamical system around $%
t_{0}=0$, we introduce the following intuitive and simple mathematical picture. Let's assume first that the deviation vector $\xi$ satisfies the condition $%
\left| \left| \xi \left( t\right) \right| \right| <t^{2}$, $t\approx 0^{+}$ \cite{rev, Ha4a}. If this is the case,  then it turns out that all the trajectories of the dynamical system are focusing together, and converge towards the origin. Let's assume now that the deviation vector $\xi$ has the property that the condition
$\left| \left| \xi \left( t\right) \right| \right| >t^{2}$, $t\approx 0^{+}$ is satisfied. In this case,  it follows that all the solutions of the system of differential equations (\ref{EM}) have at infinity a dispersing behavior \cite{rev, An00, Sa05,Sa05a}.

The dispersing/focusing behavior of the solutions of a given system of second order ordinary differential equations
can be also characterized, from the geometrical perspective introduced by the KCC theory, by considering the
algebraic properties of the deviation curvature tensor $P_i^j$. This can be done in the following way. For $t\approx 0^{+}$, the trajectories of the system of strongly nonlinear equations Eqs.~(\ref{EM}) are
bunching/focusing together if and only if the real parts of the characteristic values (eigenvalues) of the deviation tensor $P_{j}^{i}\left( 0\right) $
are strictly negative. Otherwise, the trajectories of the dynamical system have a dispersing behavior if and only if the real parts of
the characteristic values (eigenvalues) of $P_{j}^{i}\left( 0\right) $ are
strictly positive \cite{rev, An00, Sa05,Sa05a}.

By taking into account the qualitative discussion presented above, we present now the rigorous
mathematical definition of the notion of Jacobi stability for an arbitrary dynamical system, described by a system of ordinary second order differential equations, which is given by the following \cite{rev, An00,Sa05,Sa05a}:

\textbf{Definition:} \textit{Let us consider that the general system of differential equations Eqs.~(\ref{EM}), describing the time evolution of a dynamical system,  satisfies the
initial conditions}
\begin{equation*}
\left| \left| x^{i}\left( t_{0}\right) -\tilde{x}^{i}\left( t_{0}\right)
\right| \right| =0, \left| \left| \dot{x}^{i}\left( t_{0}\right) -\dot{%
\tilde{x}}^{i}\left( t_{0}\right) \right| \right| \neq 0,
\end{equation*}
\textit{defined with respect to the norm $%
\left| \left| .\right| \right| $ induced by a positive definite inner
product.}

\textit{If these conditions are satisfied, the trajectories of the dynamical system, given by Eqs.~(\ref%
{EM}), are designated as Jacobi stable if and only if the real parts of the
characteristic values (eigenvalues) of the curvature deviation tensor $P_{j}^{i}$ are everywhere strictly
negative.}

\textit{On the other hand, if the real parts of the characteristic values (eigenvalues) of the curvature deviation
tensor $P_{j}^{i}$ are strictly positive everywhere, the trajectories of the
dynamical system are designated as unstable in the Jacobi sense.}

This definition allows us to straightforwardly investigate the stability of the systems of second order differential equations, and of the associated
dynamical systems, as an alternative to the standard Lyapunov linear
stability method.

\subsection{The correlation between Lyapunov and Jacobi stability for a two
dimensional autonomous differential system}

Let us recall that the Lyapunov stability is determined by the nature and
sign of the eigenvalues of the Jacobian matrix evaluated at an equilibrium point
(fixed point). On the other hand, the Jacobi stability is given by the sign of the real part
of the eigenvalues of the curvature deviation tensor $P^i_{j}$, calculated at the
same point.



The Jacobi stability of the two dimensional systems of first order differential equations was explored
in \cite{Sa05a, Sa05}, where the authors considered a system of two arbitrary differential equations written in the general form,
\begin{equation}  \label{4n}
\frac{du}{dt}=f(u,v),\qquad \frac{dv}{dt}=g(u,v).
\end{equation}
Moreover, it was assumed that the point $(0,0)$ is a fixed point of the system (\ref{4n}), that is, $%
f(0,0)=g(0,0)=0$. If the equilibrium point is $(u_0,v_0)\neq(0,0)$, with the
change of the variable $\tilde{u}=u-u_0$, and $\tilde{v}=v-v_0$, respectively, the equilibrium
point $(u_0,v_0)$ is moved to the origin $(0,0)$.

In the approach pioneered in \cite{Sa05a, Sa05}, after relabelling the variables by denoting $v$ as $x$%
, and $g(u,v)$ as $y$, and by also assuming that the condition $g_u|_{(0,0)}\neq 0$ is satisfied by the function $g(u,v)$, it turns out
that it ism possible to eliminate from the system of the two equations (\ref{4n}) the variable $u$. By taking into account that  the point $(u,v)=(0,0)$ is a fixed
point, from the Theorem of the Implicit Functions it follows that in the
neighbourhood of the point $(x,y)=(0,0)$, the algebraic equation $g(u,x)-y=0$ has a unique solution $u=u(x,y)$
. By taking into account that $\ddot x = \dot g = g_u \, f + g_v \, y$, where the subscripts denote the partial derivatives with respect to $u$ and $v$, respectively, one obtains finally an autonomous
one-dimensional second order equation, equivalent mathematically to
the system~(\ref{4n}), and which is obtained in the general form as,
\begin{equation}  \label{5n}
\ddot x^1 + g^1(x,y) = 0,
\end{equation}
where
\begin{equation}
g^1(x,y)=-g_u(u(x,y),x) \, f(u(x,y),x) - g_v(u(x,y),x) \, y.
\end{equation}

Hence, the Jacobi stability properties of a system of two arbitrary first order differential equations can be studied in
detail via the equivalent  Eq.~(\ref{5n}) by using the methods of the KCC theory \cite{Sa05a,Sa05}. Thereupon, the KCC
stability properties of a first order system of differential equations of the form (\ref{4n}) can also be
determined easily, thus allowing an in depth comparison between the Jacobi and Lyapunov
stability properties of the two dimensional dynamical systems, which can be
performed in a straightforward manner~\cite{rev}.

Let us recall some results from~\cite{rev} which are applied in this paper.
The Jacobian matrix of~(\ref{4n}) is
\begin{equation}
J(u,v)=
\begin{pmatrix}
f_u & f_v \\
g_u & g_v%
\end{pmatrix}%
.
\end{equation}

The characteristic equation is given by
\begin{equation}  \label{vp}
\lambda^2 - (\mbox{tr} A) \lambda + \det A=0,
\end{equation}
where $\mbox{tr} A=f_u+g_v$, and $\det A=f_ug_v-g_uf_v$ are the trace and the
determinant of the Jacobian matrix $A=J|_{(0,0)}$, respectively.

The signs of the discriminant $\Delta=(f_u-g_v)^2+4f_vg_u$, and the trace and
the determinant of $A$ gives the Lyapunov (linear) stability of the fixed point $%
(0,0)$.

The system~(\ref{4n}) is equivalent from a mathematical point of view with the second order differential
equation~(\ref{5n}). Performing the Jacobi stability analysis for this last equation
in~\cite{rev}, assuming that $g_u|_{(0,0)}\neq 0$, the authors obtained the result that
the curvature deviation tensor $P^1_1$ at the fixed point $(0,0)$ is given
by
\begin{equation}
4 P^1_1|_{(0,0)} = -4g^1_{,1}|_{(0,0)}+(g^1_{;1})^2|_{(0,0)}=\Delta ,
\end{equation}
where
\begin{equation*}
\Delta:=(\mbox{tr} A)^2-4 \det A ,
\end{equation*}
is the discriminant of the characteristic equation~(\ref{vp}) and $%
A=J|_{(0,0)}$. Therefore, the authors of \cite{rev} concluded their analysis by formulating the following

\textbf{Theorem} \cite{rev} \textit{Let us consider the system of two first order ordinary differential equations, given by Eqs.~(\ref{4n}), with the fixed
point $P(0,0)$, such that $g_u|(0,0)\neq 0$. Then, the trajectory $v=v(t)$ is
Jacobi stable if and only if $\Delta<0$.}

Boehmer \textit{et. al} \cite{rev} emphasized also that if $f_v|_{(0,0)} \neq 0$,
eliminating the variable $v$ and relabeling $u$ as $x$, we get a similar
result, that the trajectory $u=u(t)$ is Jacobi stable if and only if $%
\Delta<0$.

In other words, they demonstrated that: \textit{if one consider the system of ordinary differential equations ~(\ref%
{4n}) with the fixed point $P(0,0)$ located in the origin of the coordinate system,  and satisfying the condition $g_u|(0,0)\neq0$, then the
Jacobian matrix $J$ evaluated at the point $P$ has complex proper values (eigenvalues) if and only if
$P$ satisfies the condition of being a stable point in the Jacobi sense.}

Let us recall that the condition of stability in the Lyapunov sense for a solution of the dynamical system~(\ref%
{4n})is that $\mathrm{Re} \lambda <0$ for
all roots of the characteristic (eigenvalue) equation~(\ref{vp}). The condition for the
Jacobi stability thus requires that the discriminant of the same characteristic (eigenvalue)
equation to take negative values. Consequently, Lyapunov stability is in general not
equivalent with Jacobi stability, and it worth to find out the equilibrium
points where a system is stable in both Jacobi and Lyapunov sense.

In the
next Section we will investigate the relation between Jacobi and Lyapunov
stability of the circular orbits around a GMGHS black hole.

\section{Black hole solutions in dilaton gravity}\label{sect2}

Theoretical models based on the field equations obtained from the string-like action \cite{GM88,GHS91}
\begin{equation}  \label{1a}
S=-\frac{1}{2}\int d^{4}x\sqrt{-g}\left[ R-2\gamma \left( \nabla \phi
\right) ^{2}+e^{-2\alpha \phi }F^{2}\right],
\end{equation}
where $\alpha =\mathrm{constant}>0$, and $\gamma =\mathrm{constant}>0$, have
been extensively investigated in the physical literature. In Eq.~(\ref{1a}) $%
\phi $ denotes the dilaton field, while $F$ represents the Maxwell two-form field, having the
components defined as $F_{\mu \nu }=\partial _{\mu }A_{\nu }-\partial _{\nu }A_{\mu }$.
The action (\ref{1a}) can be obtained from the string frame low energy
effective action, given by,
\begin{equation}  \label{act}
\hat{S}=-\int d^{4}x\sqrt{-\hat{g}}e^{-2\alpha \phi }\left[ \hat{R}-2\gamma
\left( \hat{\nabla}\phi \right) ^{2}+F^{2}\right] ,
\end{equation}
by using the conformal transformation $\hat{g}_{\mu \nu }=e^{2\alpha \phi
}g_{\mu \nu }$. As particular cases, the models constructed from the action (\ref{act}) include the Einstein-Maxwell theory, corresponding to $%
\alpha =\gamma =0$, and low energy string theory, which is obtained by taking $\alpha=2$, and $\gamma=2$, respectively.

The gravitational field equations derived from the action (\ref{act}) by varying the metric are given by \cite%
{GHS91}
\begin{equation}
\nabla_{\mu}\left(e^{-2\phi}F^{\mu \nu}\right)=0, \nabla ^2 \phi+\frac{1}{2}%
e^{-2\phi}F^2=0,
\end{equation}
and
\begin{equation}
R_{\mu \nu}=2\nabla _\mu \phi \nabla _\nu \phi +2e^{-2\phi}F_{\mu
\rho}F^{\rho}_\nu -\frac{1}{2}g_{\mu \nu}e^{-2\phi}F^2,
\end{equation}
respectively. By assuming for the static, spherically symmetric metric an
ansatz of the form,
\begin{equation}
ds^2=-\lambda ^2dt^2+\frac{dr^2}{\lambda ^2}+R^2d\Omega,
\end{equation}
where $\lambda$ and $R$ are functions of $r$ only, Gibbons and Maeda~\cite%
{GM88}, and, independently, three years later, Garfinkle, Horowitz and
Strominger~\cite{GHS91} found that the unique static charged black hole solution corresponding to
the action~(\ref{1a}) is given by,
\begin{equation}  \label{metr}
ds^2=-\left(1-\frac{2M}{r}\right) dt^2 +\frac{1}{\left(1-\frac{2M}{r}
	\right)} d r^2  + r \left(r-\frac{Q^2}{M}\right) (d \theta^2 + \sin^2 \theta \, d
\varphi^2)
\end{equation}
where the electric field strength, and the dilaton field are given by
\begin{equation}
F_{rt}=\frac{Q}{r^2}, \qquad e^{2\alpha \phi}=1-\frac{Q^2}{Mr}.
\end{equation}

This form of the metric is different from the usual standard form of a
spherically symmetric metric, like, for example, the Schwarzschild metric.
The above black hole solution~(\ref{metr}) is known as GMGHS black hole. For
$Q^2 < 2 M^2$, the black hole has an event horizon, and if $Q^2 = 2 M^2$, the
solution corresponds, from a physical point of view, to a naked singularity. This later case is called the extremal
GMGHS black hole solution.

An external observer cannot see the region inside the event horizons. The
main physical characteristics of a black hole can be acquired by investigating the behavior of
matter and light outside the event horizons. A massive test particle moves along
time like geodesics, and  photons move on null geodesics. The geodesics of
a GMGHS black hole were studied extensively, for both the extremal and
the non-extremal cases \cite{BB98, F12, P12, ov13, B13, B15}. In what follows we
will consider the Jacobi stability of circular time like geodesics around a
GMGHS black hole.  The existence and the Lyapunov stability of the circular orbits around GMGHS black holes was already investigated in~%
\cite{B13}.

\subsection{Geodesic equations in a static-charged black hole geometry}

We derive now the geodesics equations in the GMGHS space-time by using the mathematical formalism based on the
Euler-Lagrange equations~(\ref{EL}). The Lagrangian for the metric (\ref%
{metr}) is:
\begin{equation}
2 \mathcal{L} =-\left(1-\frac{2M}{r}\right) \dot{t}^2 + \frac{\dot{r}^2}{%
\left(1-\frac{2M}{r} \right)} + r \left(r-\frac{Q^2}{M}\right) \left(\dot{\theta}^2 + \sin^2 \theta \, \dot{\varphi}^2 \right)  \label{lagr}
\end{equation}
where by a dot we have denoted the differentiation with respect to $\tau$ - an affine
parameter defined along the geodesic line. The parameter $\tau$ is chosen so that the Lagrangian $\mathcal{L}$ satisfies the condition $2 \mathcal{L}%
=-1$ on a time-like geodesics, $2 \mathcal{L}=0$ on a null geodesics, and $2
\mathcal{L}=1$ on a space-like geodesics, respectively. Moreover, all the functions $F_i$, $i=1,2,\dots n$,
from the right hand side of Eq.~(\ref{EL}) are zero.

We recall that the coordinates $t$ and $\varphi$ are cyclic. Thus, we obtain
that
\begin{equation}  \label{ien}
\left(1-\frac{2M}{r}\right) \dot{t} = \text{constant} = E ,
\end{equation}
is the energy integral, where $E$ is the total energy of the particle, and
\begin{equation}  \label{imc}
2 \, \sin^2\theta \, \cdot r \left(1-\frac{2M}{r}\right) \dot{\varphi}={\rm constant}=L,
\end{equation}
is the integral of the angular momentum.

The Euler-Lagrange equation for $\theta$ is
\begin{equation}  \label{ecteta}
\frac{d}{d \tau}\left[ r \left( r - \frac{Q^2}{M} \right) \dot{\theta} %
\right] = r \left( r - \frac{Q^2}{M} \right) \sin \theta \cos \theta \cdot
\dot{\varphi}^2\,.
\end{equation}
We note that if $\theta = \pi/2$, then $\dot{\theta}\equiv 0$, $\ddot{\theta}
\equiv 0$, and therefore $\theta=\pi/2$ is located on the geodesic curve. Thus, the motion of a massive particle is planar. On the other hand, the
angular momentum integral (\ref{imc}) takes the form,
\begin{equation}
r \left(1-\frac{2M}{r}\right) \dot{\varphi} = L
\end{equation}
where $L$ represents, from a physical point of view, the angular momentum of the particle,  oriented in the direction of an axis perpendicular to the plane in which
the motion of the particle takes place.

Being complicated, the Euler-Lagrange equation for $r$ is replaced with the
constancy of the Lagrangian,
\begin{equation}  \label{et}
\frac{1}{2}\left( \frac{dr}{d \tau} \right)^2 +\frac{1}{2} \left( 1 - \frac{2 M}{r} \right) \left[%
\frac{L^2}{r \left( r - \frac{Q^2}{M} \right)} - \epsilon \right] = E^2
\end{equation}
where the constant $\epsilon$ takes the numerical values $\epsilon=-1$ for time like geodesics, $\epsilon=0$ for null geodesics,
and $\epsilon=+1$ for space like geodesics, respectively. The second term in the left-hand
side of Eq.~(\ref{et}) is the effective potential. Hence, Eq.~(\ref{et} can be written as
\begin{equation}
\frac{1}{2}\left( \frac{dr}{d \tau} \right)^2 + V(r) = E^2
\end{equation}

A massive test particle moving freely around a GMGHS black hole describes a time like
geodesic line. For them $\epsilon=-1$, and the effective potential becomes
\begin{equation}  \label{Veff}
V(r) =\frac{1}{2} \left( 1 - \frac{2 M}{r} \right) \left[\frac{L^2}{r \left( r - \frac{%
Q^2}{M} \right)} + 1 \right]\,.
\end{equation}

\begin{figure*}[htbp]
\centering
\includegraphics[scale=0.53]{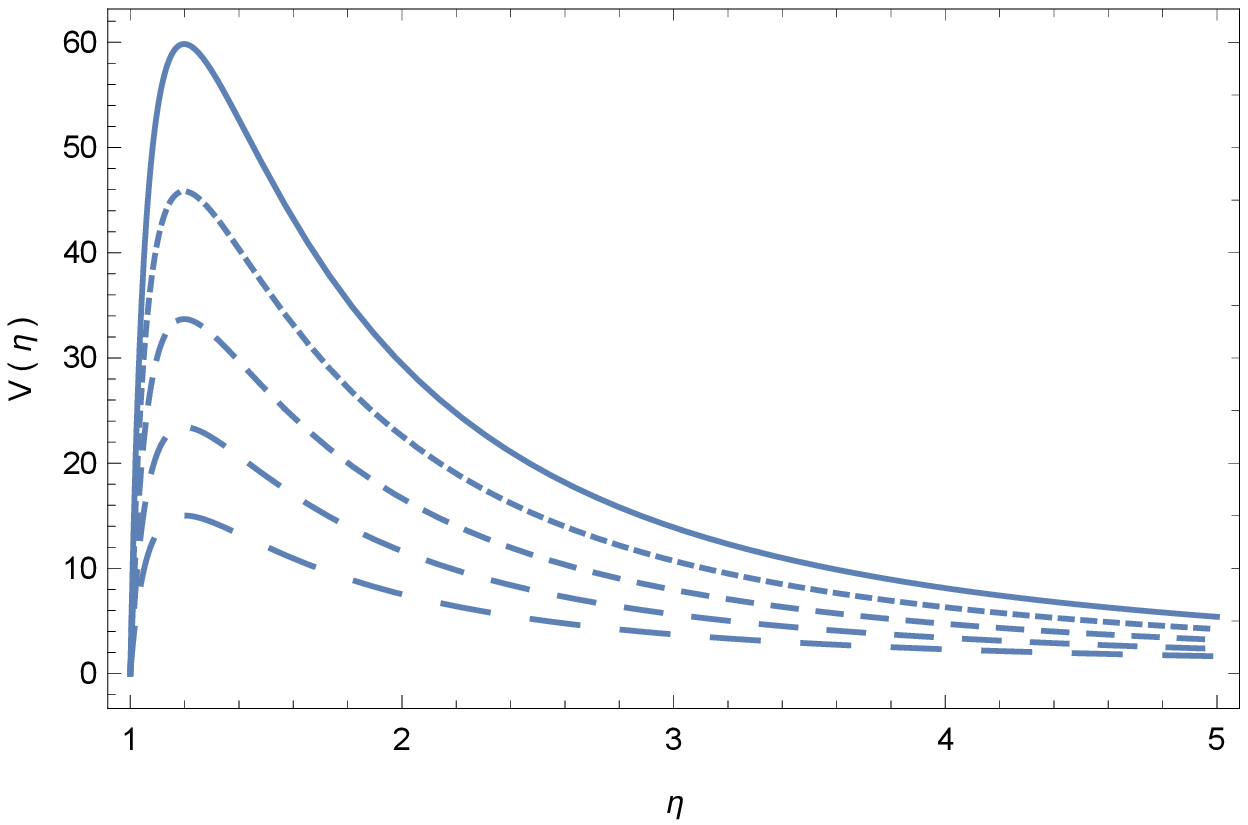}
\includegraphics[scale=0.53]{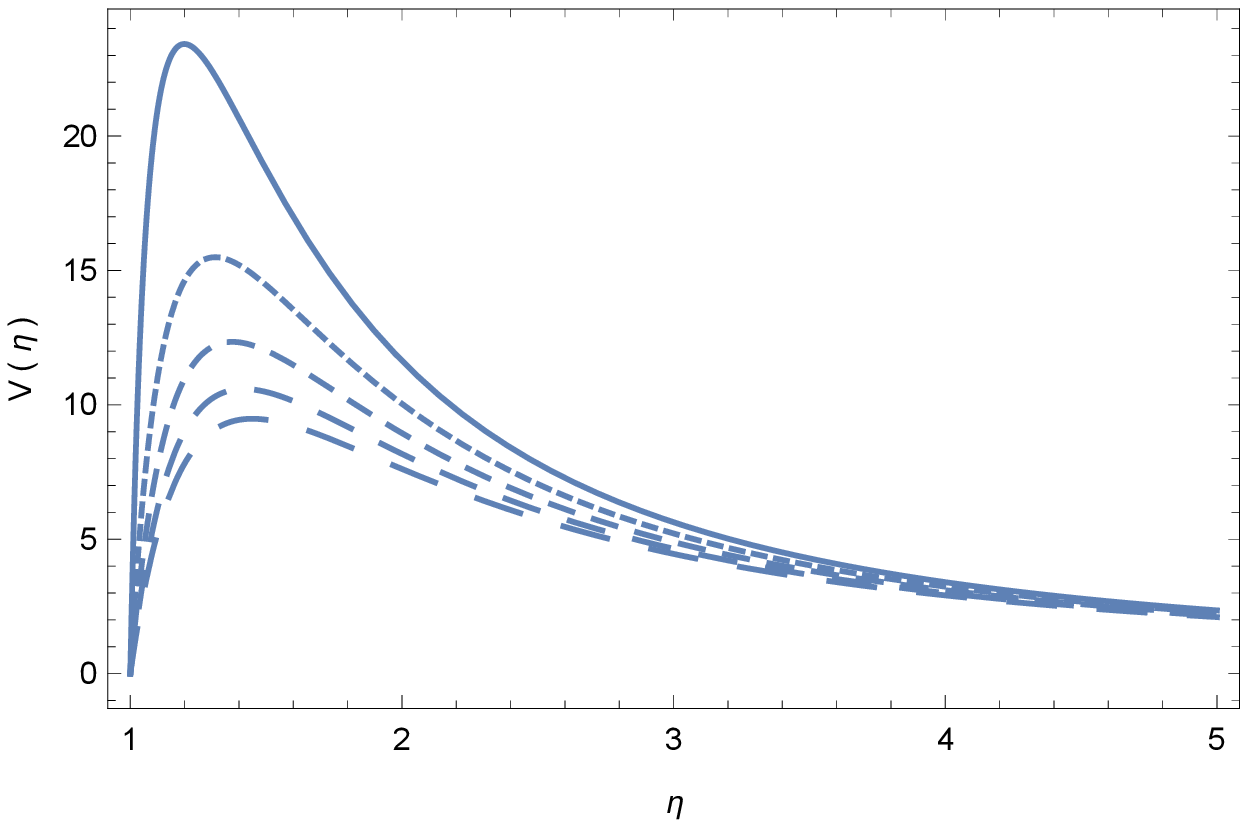}
\caption{Variation as a function of $\eta$ of the effective
potential $V(\eta)$ of the GMGHS black hole for $q=0.95$, and $l=16$
(solid curve), $l=14$ (dotted curve), $l=12$ (short dashed curve), $l=10$
(dashed curve), and $l=8$ (long dashed curve), respectively (left panel),
and for $l=10$, and $q=0.95$ (solid curve), $q=0.85$ (dotted curve), $q=0.75$
(short dashed curve), $q=0.65$ (dashed curve), and $q=0.55$ (long dashed
curve, respectively (right panel).}
\label{Fig1}
\end{figure*}

By introducing the new variable $\eta$, defined as $r=2M\eta$, the effective
potential (\ref{Veff}) becomes
\begin{equation}
V(\eta)=\frac{1}{2}\left(1-\frac{1}{\eta}\right)\left[\frac{l^2}{\eta^2\left(1-\frac{q^2%
}{\eta}\right)}+1\right],
\end{equation}
where we have denoted $l^2=L^2/4M^2$, and $q^2=Q^2/2M^2$, respectively. The
variation of the potential $V(\eta)$ is presented, for different values of $%
l $ and $q$, in Fig.~\ref{Fig1}.

The first derivative of the potential can be obtained immediately as
\begin{equation}
V^{\prime }(\eta )=\frac{1}{2\eta ^{2}}\left[ 1+\frac{l^{2}}{\eta ^{2}}\frac{%
(1-\frac{2}{\eta })q^{2}-2(1-\frac{3}{2\eta })\eta }{\left( 1-\frac{q^{2}}{%
\eta }\right) ^{2}}\right] .
\end{equation}

\begin{figure*}[htbp]
\centering
\includegraphics[scale=0.53]{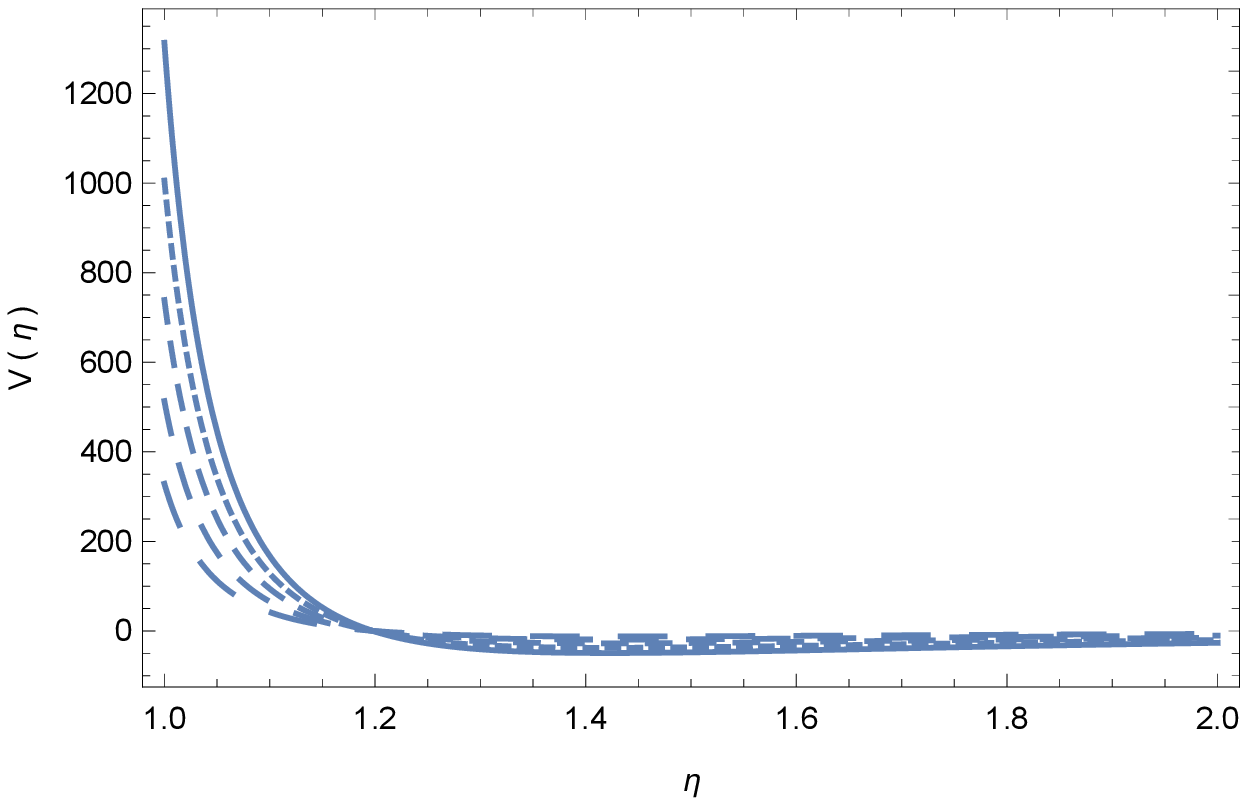}
\includegraphics[scale=0.53]{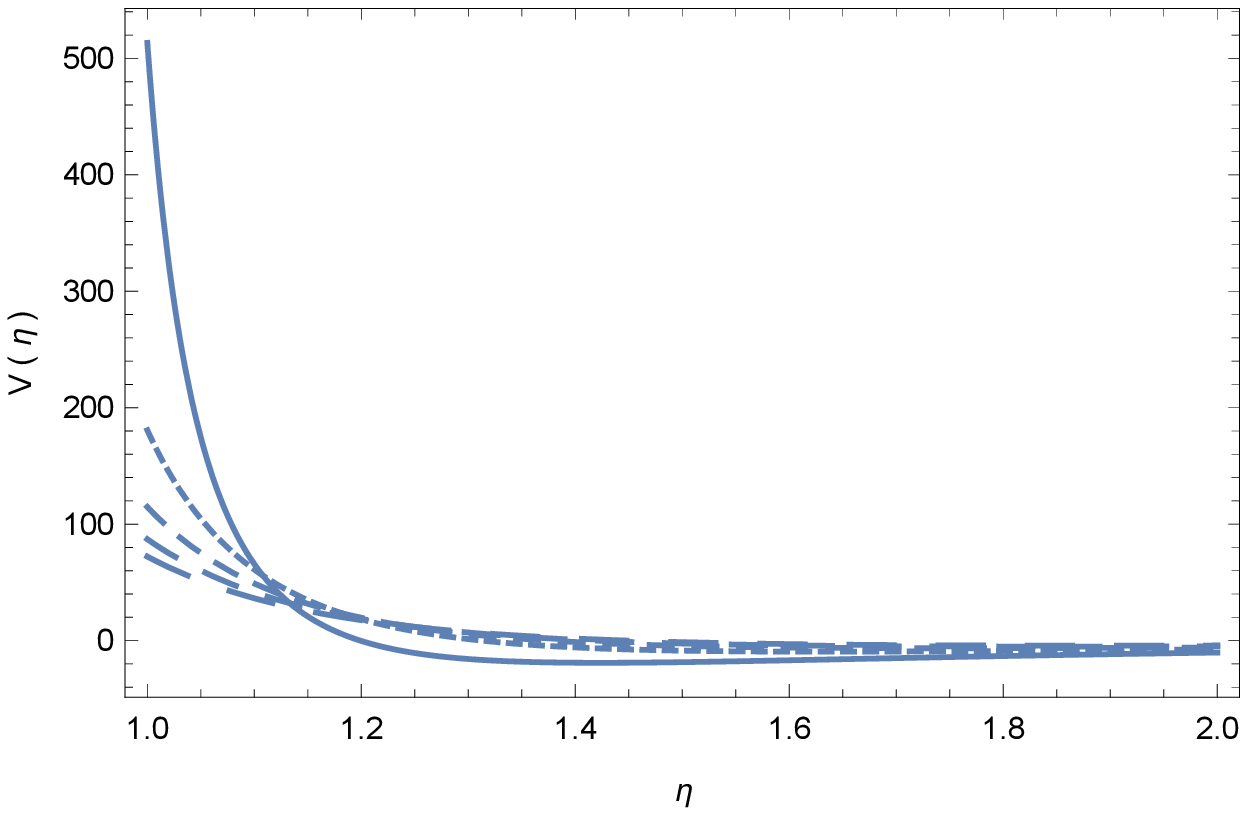}
\caption{Variation as a function of $\eta$ of the derivative $%
V^{\prime }(\eta)$ of the effective potential of the GMGHS black
hole for $q=0.95$, and $l=16$ (solid curve), $l=14$ (dotted curve), $l=12$
(short dashed curve), $l=10$ (dashed curve), and $l=8$ (long dashed curve),
respectively (left panel), and for $l=10$, and $q=0.95$ (solid curve), $%
q=0.85$ (dotted curve), $q=0.75$ (short dashed curve), $q=0.65$ (dashed
curve), and $q=0.55$ (long dashed curve, respectively (right panel).}
\label{Fig2}
\end{figure*}

The variation of the derivative $V^{\prime }(\eta)$ of the potential as a
function of $\eta$ is represented in Fig.~\ref{Fig2}.

For the second derivative of the potential we obtain
\begin{equation}
V^{\prime \prime }(\eta )=-\frac{2}{\eta ^{3}}\left[ 1-\frac{l^{2}}{\eta ^{4}%
}\frac{3(1-\frac{2}{\eta })\eta ^{2}+(1-\frac{3}{\eta })q^{4}-3(1-\frac{8}{%
3\eta })\eta q^{2}}{\left( 1-\frac{q^{2}}{\eta }\right) ^{3}}\right] .
\end{equation}

\begin{figure*}[htbp]
\centering
\includegraphics[scale=0.53]{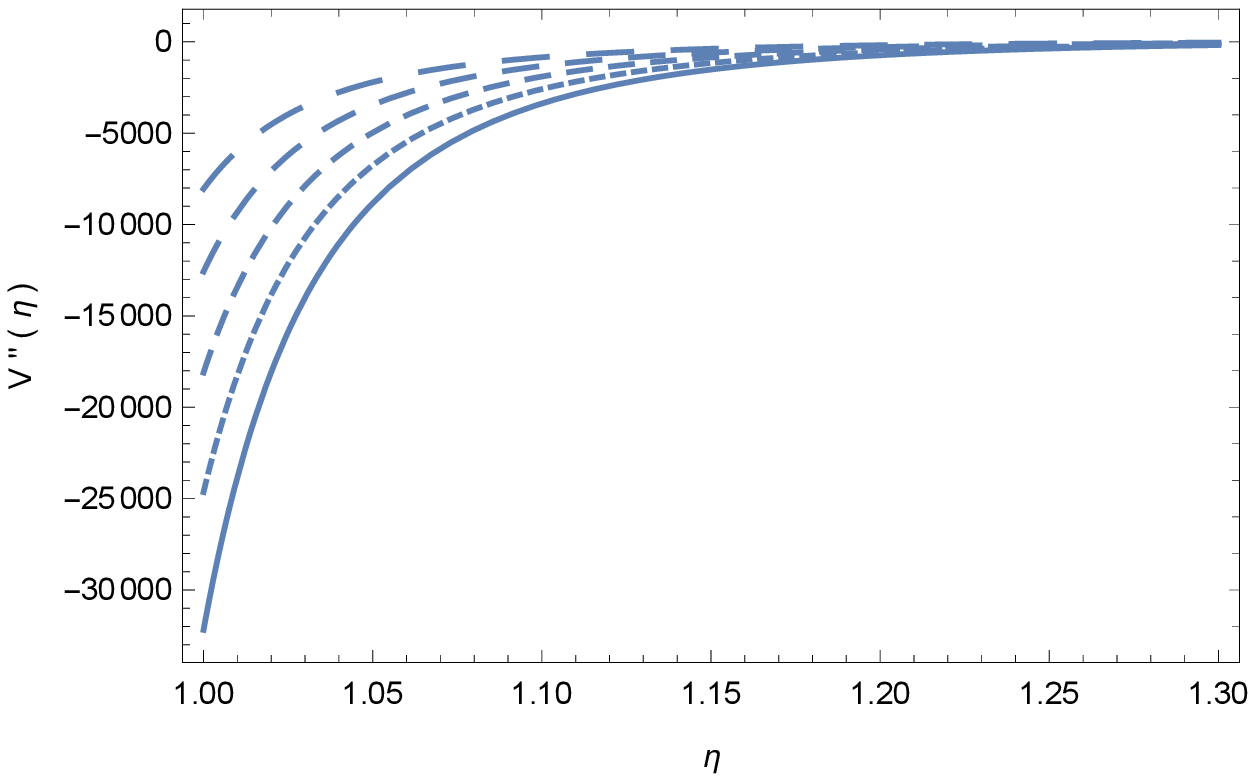}
\includegraphics[scale=0.53]{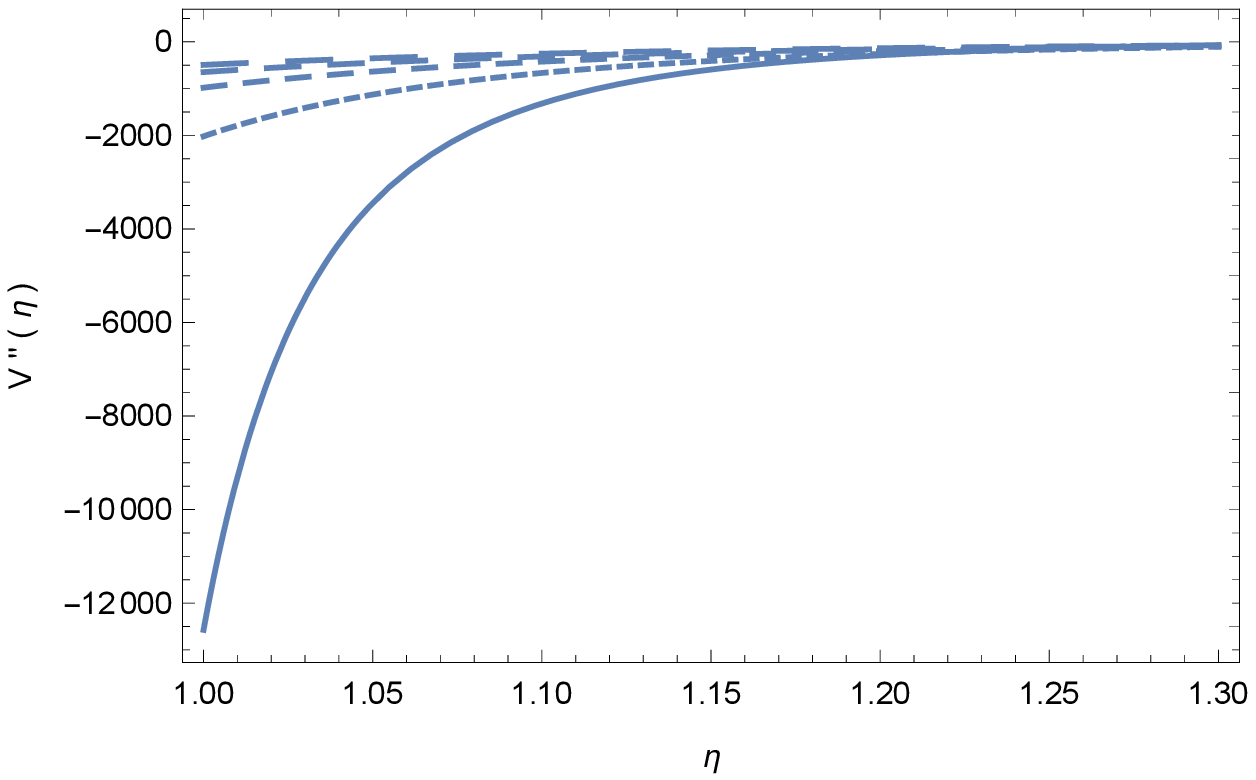}
\caption{Variation as a function of $\eta$ of the second derivative $%
V^{\prime \prime}(\eta)$ of the effective potential of the GMGHS black
hole for $q=0.95$, and $l=16$ (solid curve), $l=14$ (dotted curve), $l=12$
(short dashed curve), $l=10$ (dashed curve), and $l=8$ (long dashed curve),
respectively (left panel), and for $l=10$, and $q=0.95$ (solid curve), $%
q=0.85$ (dotted curve), $q=0.75$ (short dashed curve), $q=0.65$ (dashed
curve), and $q=0.55$ (long dashed curve, respectively (right panel).}
\label{Fig3}
\end{figure*}

The variation of $V''(\eta)$ is presented, for different values of $l$ and $q$, in Fig.~\ref{Fig3}.

\begin{figure}[htbp]
\centering
\includegraphics[scale=0.67]{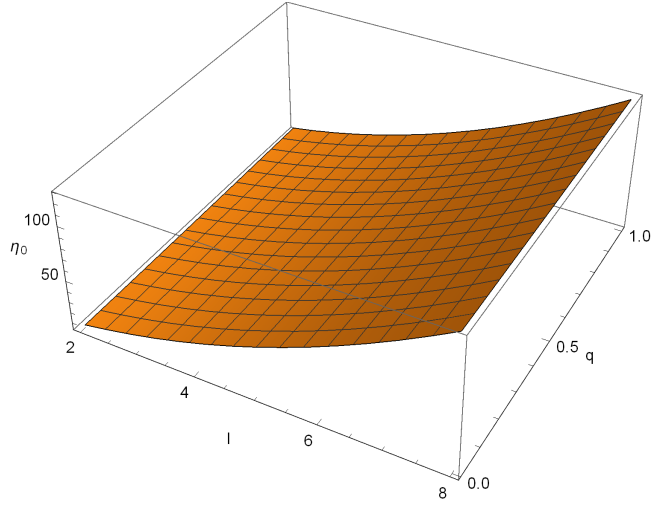}
\caption{Variation as a function of $l^2$ and $q^2$ of the solution $\eta_0$ of the algebraic equation  $V'(\eta_0)=0$.}
\label{Fig4}
\end{figure}

The dependence of the real solution $\eta_0$  of the equation $V'(\eta_0)=0$ on the parameters $l^2$ and $q^2$ of the GMGHS black hole solution is represented in Fig.~\ref{Fig4}.

%

\section{Stability of the circular orbits of the free test particles in a GMGHS
spacetime}
\label{sect3}

Next we will consider the stability in Lyapunov and Jacobi sense of the
circular orbits on which massive test particle move freely around a GMGHS black hole.

\subsection{Lyapunov stability analysis}

 Eq.~(\ref{et}) is the starting point in the dynamical systems approach for the analysis of the stability of the geodesic curves in the GMGHS geometry.
Differentiating Eq.~(\ref{et}) with respect to $\tau$ and dividing the
result with $\dot{r}$, we obtain the following second order differential
equation
\begin{equation}  \label{2de}
\ddot{r}=-\frac{d V}{dr}.
\end{equation}

The Eq.~(\ref{2de}) corresponds to following system of first order
differential equations
\begin{equation}  \label{La}
\frac{d r}{d \tau}=p\,,\qquad \frac{d p}{d \tau}=-\frac{d V}{d r}\,.
\end{equation}

The Jacobian matrix of the system~(\ref{La}) is
\begin{equation}  \label{J}
J=%
\begin{pmatrix}
0 & 1 \\
-V^{\prime \prime }(r) & 0%
\end{pmatrix}
\,,
\end{equation}
where $^{\prime }$ means differentiation with respect to $r$.

The characteristic equation is
\begin{equation}  \label{ec}
\lambda^2 + V^{\prime \prime }(r)=0,
\end{equation}
and the proper values (eigenvalues) of the Jacobian matrix associated to the system~(\ref{La}) are given by
\begin{equation}  \label{l}
\lambda=\pm\sqrt{-V^{\prime \prime }(r)},
\end{equation}
and so a simple fixed point $(r_0,0)$ of~(\ref{La}) is a saddle point if $%
(V^{\prime \prime }(r_0)<0)$ and a center if $(V^{\prime \prime }(r_0)>0)$.

In~\cite{B13} the existence and stability in the sense of Liapunov of
circular timelike geodesics around a GMGHS black hole was explored. The
study showed that for certain values of the parameters, outside the black
hole, there are two circular geodesics. In other words, outside the black
hole, we can find two values $r_0>2M$ so that $V^{\prime }(r_0)=0$, one for
a minimum of the effective potential ($V^{\prime \prime }(r_0)<0$) and the
other for a maximum of the potential ($V^{\prime \prime }(r_0)>0$).

If $V^{\prime \prime }(r_0)<0$, the eigenvalues~(\ref{l}) of the
linearization of the system~(\ref{La}) are real and have opposite sign,
therefore the fixed point $(r_0,0)$ of the system is a saddle point, which
is Liapunov unstable. A saddle point is a hyperbolic point, and so, based on
the Hartman-Grobman theorem, the fixed point $(r_0,0)$ of the system~(\ref%
{La}) is Liapunov unstable.

If $V^{\prime \prime }(r_0)>0$, the values of $\lambda$ from~(\ref{l}) are
purely complex conjugate and the study of the linear stability of system at
the fixed point $(r_0,0)$ begins with the search of a Liapunov function for
the system~(\ref{La}). We note that the function
\begin{equation}  \label{Lf}
\mathbf{V}(r,p)=\frac{p^2}{2}+V(r)
\end{equation}
has the property that $\nabla \mathbf{V}(r,p) \cdot f(r,p)=0, \forall (r,p)$%
, where $f(r,p)=\left(p,-d V/d r\right)$, meaning that the function~(\ref{Lf}%
) could be chosen as a Liapunov function. Further, we have to check if $%
\mathbf{V}$ fulfills the condition of the Liapunov stability theorem. In
other words, we have to verify if the fixed point is a local minimum of $%
\mathbf{V}$. Therefore, we compute the Hessian matrix of $\mathbf{V}$
\begin{equation}  \label{H}
\mathcal{H}_{\mathbf{V}}=%
\begin{pmatrix}
V^{\prime \prime }(r) & 0 \\
0 & 1%
\end{pmatrix}
\,.
\end{equation}

We note that the matrix~(\ref{H}) is positive definite when $V^{\prime
\prime }(r)>0$, meaning that the fixed point $(r_0,0)$ of the linearized
system corresponding to~(\ref{La}) is a center. And thus, we finally conclude that the
circular orbits around a GMGHS black hole, $r=r_0={\rm constant}$, are Lyapunov stable when
$V^{\prime \prime }(r_0)>0$, and Lyapunov unstable when $V^{\prime \prime
}(r_0)<0$.


\subsection{Jacobi stability analysis}

In this Section we will perform first the study of the Jacobi stability analysis of Eq.~(\ref{2de}), giving the geodesic trajectories of a massive particle in the GMGHS geometry. Then, we will consider the stability of the circular orbits in both Lyapunov and KCC approaches.

\paragraph{{\bf Stability of the GMGHS orbits in the general case}.}  The affine parameter $\tau$ in the geodesic equation (\ref{2de}) is an absolute invariant, and hence all the results of the KCC theory can be applied to this case. By introducing the dimensionless radial coordinate $\eta$,  the geodesic equation of motion in the GMGHS geometry takes the form
\begin{equation}\label{Jac}
\frac{d^{2}\eta \left( \tau \right) }{d\tau ^{2}}+\frac{1}{4M^{2}}\frac{%
dV\left( \eta \right) }{d\eta }=0,
\end{equation}%
or, equivalently,
\begin{eqnarray}
&&\frac{d^{2}\eta (\tau )}{d\tau ^{2}}+\frac{1}{4M^{2}}\Bigg\{ \frac{1}{\eta
^{2}(\tau )}\left[ 1+\frac{l^{2}}{\eta ^{2}(\tau )\left( 1-\frac{q^{2}}{\eta
(\tau )}\right) }\right] \nonumber\\
&&-\left( 1-\frac{1}{\eta (\tau )}\right) \left[
\frac{2l^{2}}{\eta ^{3}(\tau )\left( 1-\frac{q^{2}}{\eta (\tau )}\right) }+%
\frac{l^{2}q^{2}}{\eta ^{4}(\tau )\left( 1-\frac{q^{2}}{\eta (\tau )}\right)
^{2}}\right] \Bigg\}=0.
\end{eqnarray}%
By denoting $\eta \left( \tau \right) =x^{1}$, and $\eta ^{\prime }\left(
\tau \right) =y^{1}$, Eq. (\ref{Jac}) takes the form
\begin{equation}
\frac{d^{2}x^{1}}{d\tau ^{2}}+2G^{1}\left( x^{1}\right) =0,
\label{GenJ}
\end{equation}%
where
\begin{equation}
G^{1}\left( x^{1}\right) =\frac{1}{8M^{2}}\frac{dV\left( x^{1}\right)
}{dx^{1}},
\end{equation}%
or,
\begin{eqnarray}
&&G^{1}\left( x^{1}\right) =\frac{1}{8M^{2}}\Bigg\{ \frac{1}{\left(
x^{1}\right) ^{2}}\left[ 1+\frac{l^{2}}{\left( x^{1}\right) ^{2}\left( 1-%
\frac{q^{2}}{x^{1}}\right) }\right] \nonumber\\
&&\left( 1-\frac{1}{x^{1}}\right) \left[
\frac{2l^{2}}{\left( x^{1}\right) ^{2}\left( 1-\frac{q^{2}}{x^{1}}\right) }+%
\frac{l^{2}q^{2}}{\left( x^{1}\right) ^{2}\left( 1-\frac{q^{2}}{x^{1}}%
\right) ^{2}}\right] \Bigg\}.
\end{eqnarray}
The nonlinear connection associated to Eq. (\ref{GenJ}) is obtained as,
\begin{equation}
N_{1}^{1}=\frac{\partial G^{1}\left(x^1\right)}{\partial y^{1}}\equiv 0.
\end{equation}

For the Berwald connection we have
\begin{equation}
G_{11}^{1}=\frac{\partial N_{1}^{1}}{\partial y^{1}}\equiv 0.
\end{equation}

For the curvature deviation tensor we obtain now the simple expression,
\begin{equation}
P_1^1=-2\frac{\partial G^1\left(x^1\right)}{\partial x^1}=-\frac{1}{4M^2}V''\left(x^1)\right).
\end{equation}

Hence, the geodesic trajectories of a massive test particle in the spherically symmetric GMGHS black hole are Jacobi stable if the condition $-\left.V''\left(x^1\right)\right|{x^1=x^1_0}<0$. On the other hand, we obtain a geometric interpretation of the second derivative of the potential, as giving the deviation curvature tensor of the geodesic trajectories. On the other hand, for the first KCC invariant of the system we obtain
\begin{equation}
\epsilon ^1=2G^1\left(x^1\right)=\frac{1}{4M^{2}}\frac{dV\left( x^{1}\right).
}{dx^{1}}
\end{equation}

Hence, the first derivative of the potential, representing physically the force acting on the particle, has a geometric interpretation as the first KCC invariant. Moreover, the geodesic deviation equations take the form
\begin{equation}
\frac{d^2\xi ^1}{d\tau ^2}+\frac{1}{4M^{2}}V''\left(x^1\right)\xi ^1=0.
\end{equation}

\begin{figure*}[htbp]
	\centering
	\includegraphics[scale=0.58]{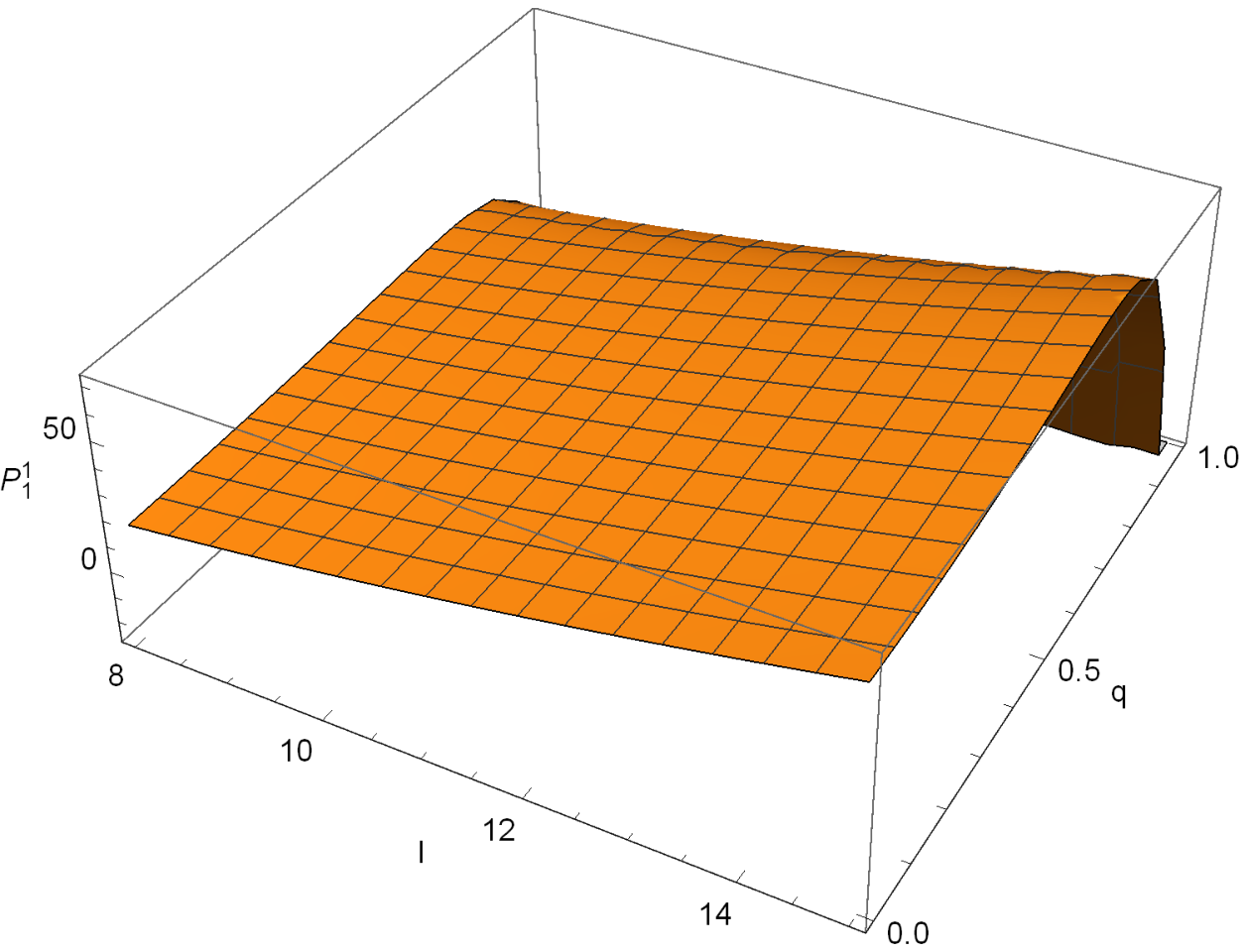}
	\includegraphics[scale=0.58]{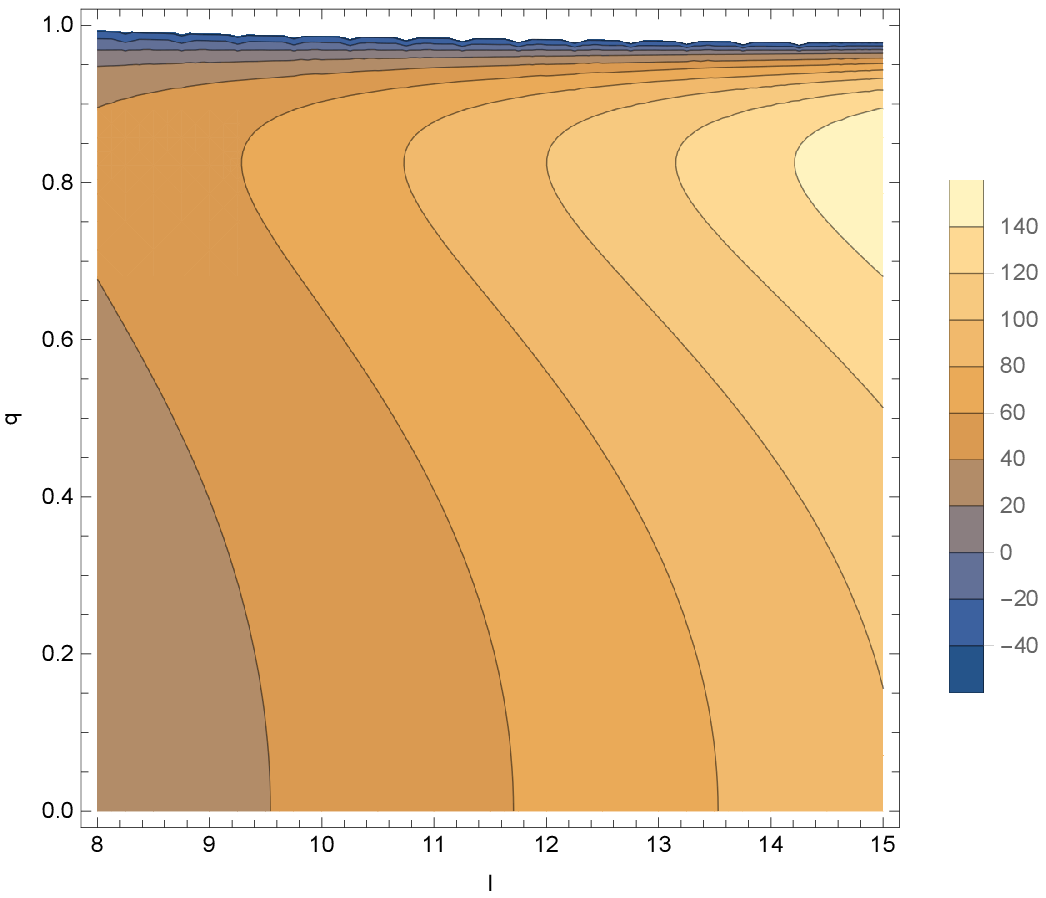}
\caption{Variation of the curvature deviation tensor $P_{1}^{1}\left(\eta, l,q\right)$ as a function of $l$ and $q$ for $\eta =1.75$ (left panel), and the contour plot of the curvature deviation tensor for $\eta=1.25$ (right panel).}
	\label{new1}
\end{figure*}

The variation of the deviation curvature tensor of the GMGHS black hole is represented, as a function of the solution parameters $l$ and $q$, for a fixed value of $\eta$, in Fig.~\ref{new1}. The contour plot of the deviation tensor is also represented.

\paragraph{{\bf Stability of the circular orbits}. }{In this Section we will perform the Jacobi stability of the system~(\ref{La}%
) using the results obtained by Boehmer \textit{et. al} in \cite{rev}. For
the system~(\ref{La}), we consider the Jacobian matrix of the system and
evaluate its trace and determinant
\begin{equation}
\mbox{tr} J=0\,, \quad\det J =V^{\prime \prime }(r) \,.
\end{equation}
Thus, the discriminant of the characteristic equation becomes
\begin{equation}
\Delta={\rm tr}(J)^2-4{\rm det}(J)=-4V^{\prime \prime }(r)\,.
\end{equation}
Based on the Theorem which makes the link between the discriminant of the
characteristic equation and the Jacobi stability, we can conclude that the
circular orbit of a free particle moving in an GMGHS spacetime, $r=r_0$, is
Jacobi stable when $V^{\prime \prime }(r_0)>0$ and Jacobi unstable for $%
V^{\prime \prime }(r_0)<0$.

We note that we have found the same condition for Jacobi stability as for the
for Lyapunov stability, meaning that for the circular orbits on which massive
test particle move around a GMGHS black hole, the two types of stability coincides.



We consider now some particular cases of stability, corresponding to some specific values of the parameters $l$ and $q$ of the GMGHS black hole. By taking $l=8$, and $q=0.95$, it turns out that the equation $V'(\eta)=0$ has two real solutions satisfying the condition $\eta >1$, given by $\eta _1=1.1996$ and $\eta_2= 127.8522$. By evaluating the second derivative of the potential in these points gives $\left.V''(\eta)\right|_{\eta =1.19996}=-185.5041<0$, and $\left.V''(\eta)\right|_{\eta =127.8522}=+2.39\times 10^{-7}>0$, respectively. Hence, we can conclude that the circular trajectory located at $r=2M\eta _1$ is both Lyapunov and Jacobi unstable (left panel in Fig.~\ref{Fig4a}), while the circular trajectory located at $r=2M\eta _2$ is both Lyapunov and Jacobi stable (right panel in Fig.~\ref{Fig4a}).

\begin{figure*}[htbp]
	\centering
	\includegraphics[scale=0.35]{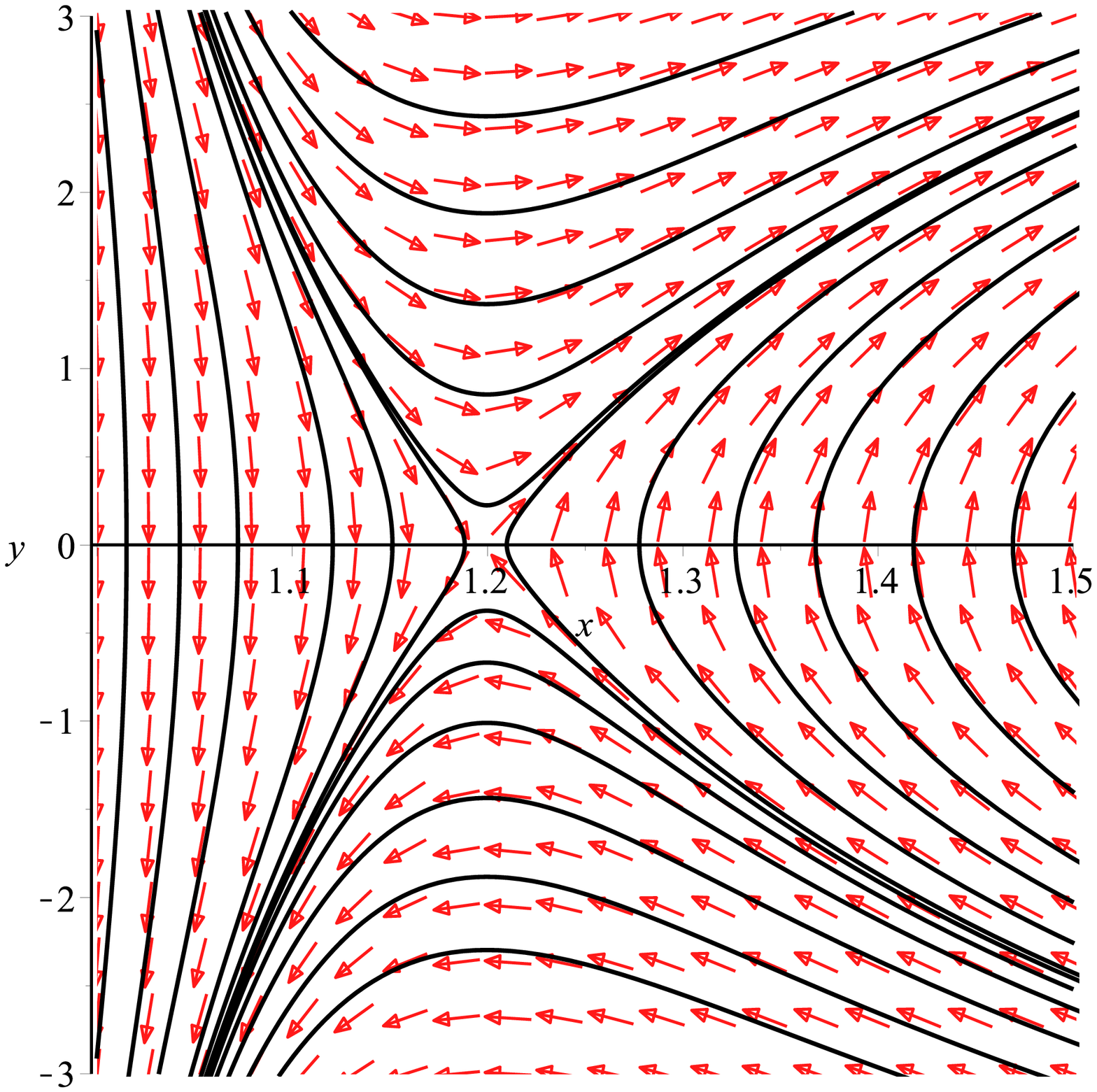}
	\includegraphics[scale=0.35]{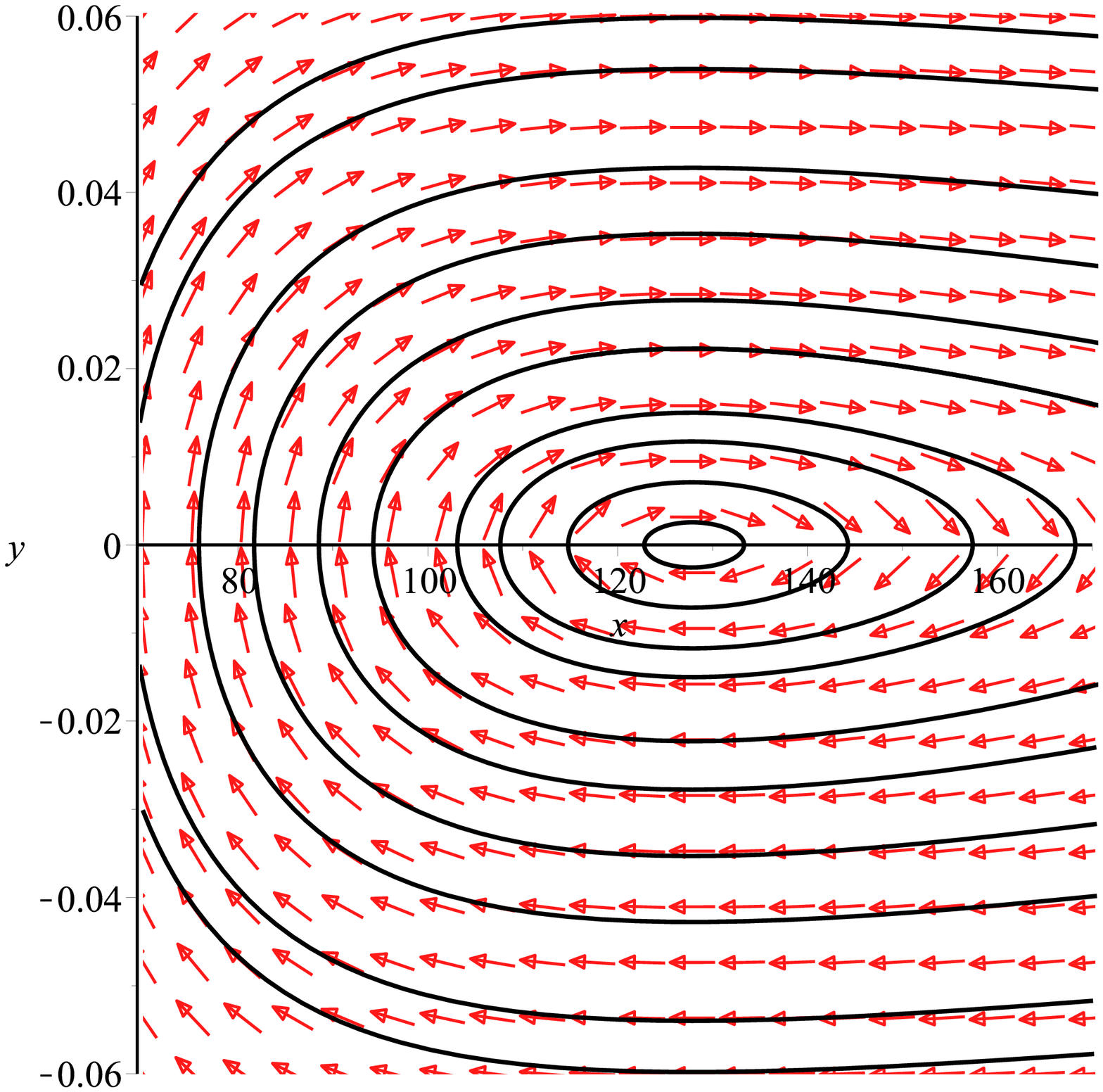}
	\caption{The phase plane for the specific values of the parameters $l=8$ and $q=0.95$. The behavior of the solution near point $\eta_1=1.1996$, represented in the left panel, shows that it is a saddle point. The behavior of the solution near the second point $\eta_2= 127.8522$, from the right panel, shows that the point is a center.}
	\label{Fig4a}
\end{figure*}

Let us now consider the Lyapunov stability corresponding to the value  $l=3/2$ of the parameter $l$ of a GMGHS black hole. By taking $q \in \{0.55, 0.6512, 0.75\}$, the equation $V'(\eta)=0$ has no real solution, two real and equal solutions, and two different real solutions, satisfying the condition $\eta >1$, meaning that outside the black hole, there are no circular orbits, one circular orbit and two circular orbits, respectively. For $q=0.6512$, $\eta _1=\eta_2= 2.5243$, by evaluating the second derivative of the potential, we get $\left.V''(\eta)\right|_{\eta =2.5243}=0$, the point $\eta =2.5243$ is an inflection point for the potential, which leads to a cusp in the phase diagram (see the middle panel in Fig.~\ref{Fig5a}). For $q=0.75$, $\eta_1=1.8333$ and $\eta_2=3.3836$ and by evaluating the second derivative of the potential in these points, we get $\left.V''(\eta)\right|_{\eta =1.8333}=-0.111<0$ and $\left.V''(\eta)\right|_{\eta =3.3836}=+0.0075>0$, respectively. Therefore, we can conclude that the circular trajectory located at $r=2M\eta _1$ is Lyapunov unstable, and the circular trajectory located at $r=2M\eta _2$ is Lyapunov stable (see the right panel in Fig.~\ref{Fig5a}).
\begin{figure*}[htbp]
	\centering
	\includegraphics[scale=0.23]{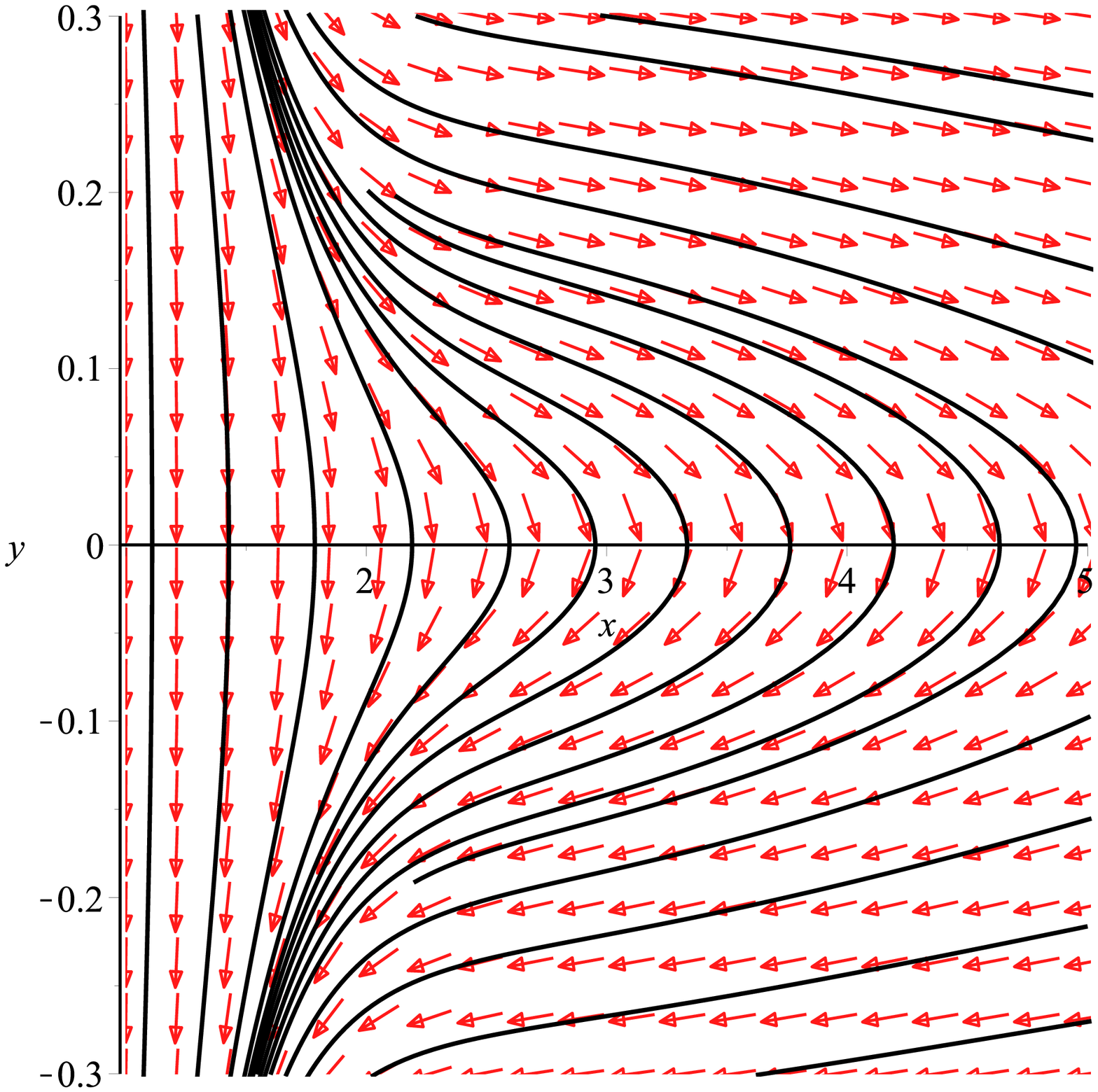}
	\includegraphics[scale=0.23]{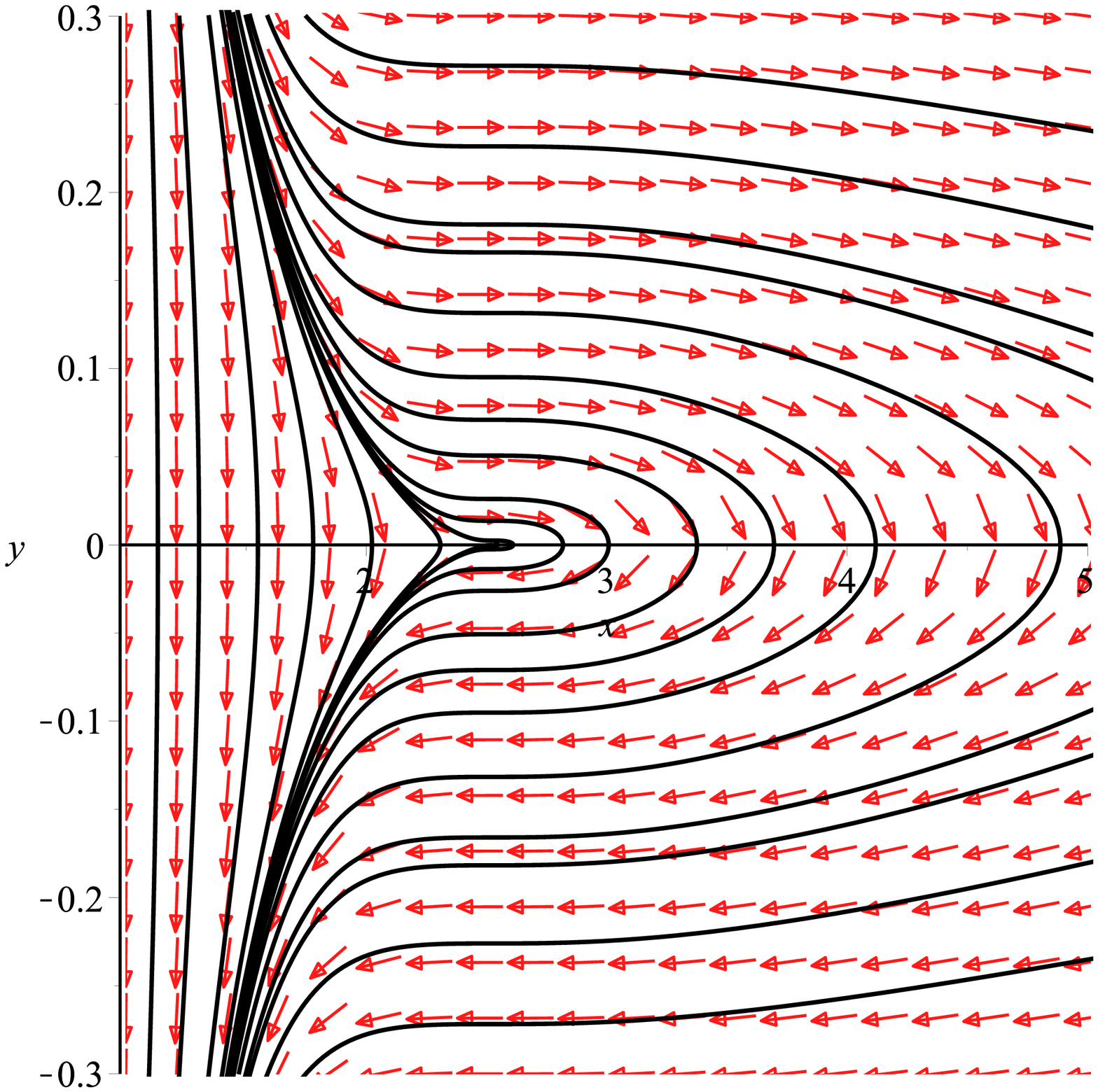}
	\includegraphics[scale=0.23]{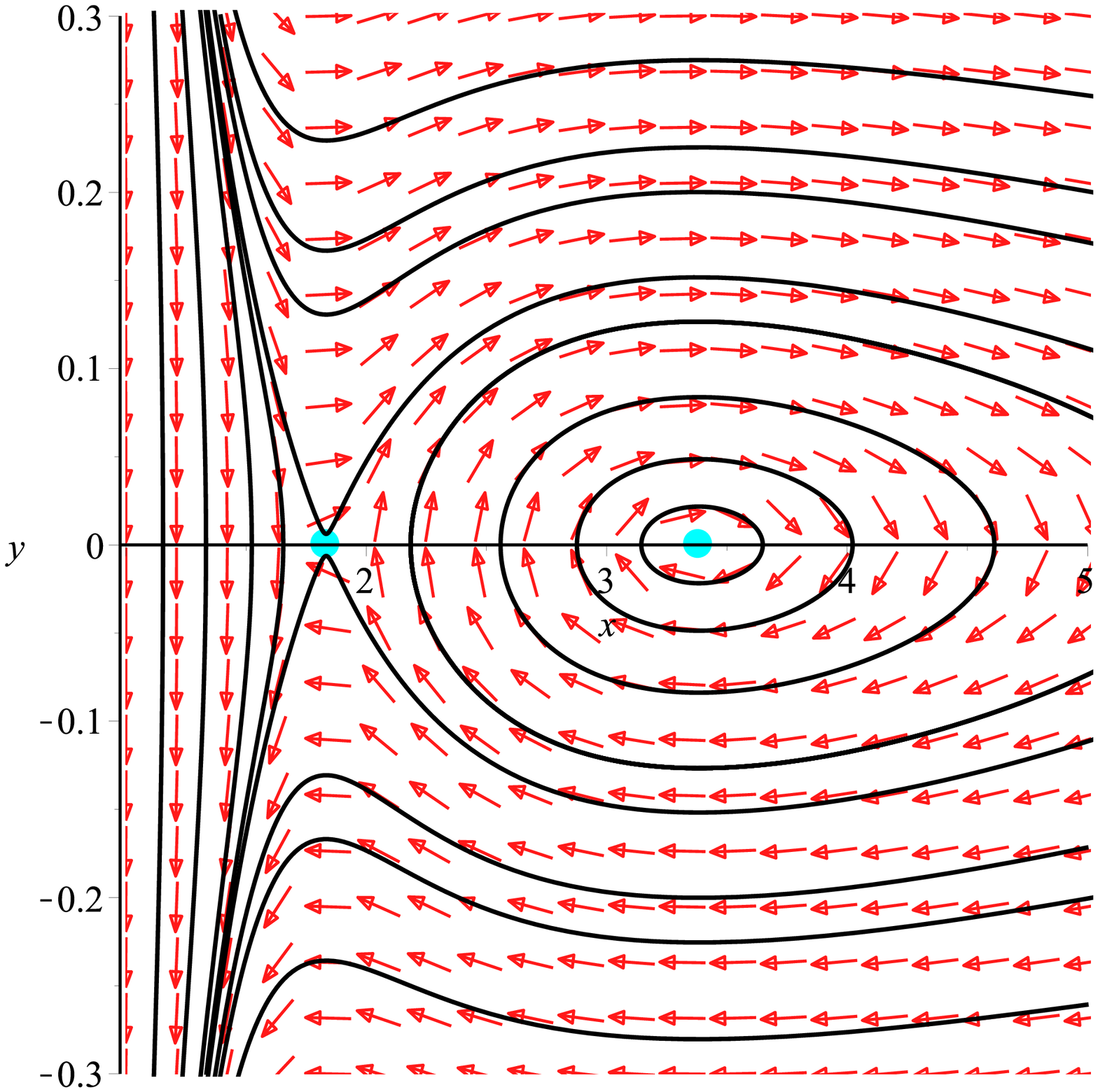}
	\caption{The phase plane for the specific values of the parameters $l=3/2$ and $q \in \{0.55, 0.6512, 0.75\}$. The left panel corresponds to $q=0.55$, in which case the first derivative of the potential has no solution for $\eta >1$, thus there are no critical points in the phase diagram. The middle panel is for $q=0.6512$, in which case the first derivative of the potential has a double solution for $\eta >1$, meaning that in the phase plane there is a cusp in the point $(2.5243,0)$. The right panel corresponds to $q=0.75$, in which case there are two circular orbits around the black hole, corresponding to $\eta_1=1.8333$ and $\eta_2=3.3836$ respectively, with the first generating a saddle point, while the second a center.}
	\label{Fig5a}
\end{figure*}

\section{Concluding remarks}\label{sect4}

In the present paper, we have first revisited, and carefully investigated, two methods of stability
analysis: Lyapunov stability, and the Jacobi stability approaches,
respectively. The Lyapunov stability analysis is done by the linearization of the system of differential equations describing
the dynamical system at the fixed points. On the other hand, and Jacobi
stability involves the perturbations of a whole set of trajectories.
Intuitively, the Jacobi stability indicates how the trajectories bunch
together, or disperse, when approaching the fixed point.

As an application of the two stability methods we have comparatively
investigated the behavior of the trajectories of the solutions for a
specific charged black hole solution that originates in the low energy limit
of string theory, called the GMGHS solution. The study of the properties of the geodesic curves in black hole geometries is an important field of investigation \cite{BB98,F12,P12,ov13,B13,B15}, which could lead not only to a better understanding of the theoretical  properties of these objects, but could also open new  perspectives on their observational detection. Moreover, black holes represent a fertile testing ground of modified gravity theories.

An analysis of the stability of the orbits in the Schwarzschild geometry was performed in \cite{X3}, by using both Lyapunov and Jacobi stability approaches. As a result of this study it was shown that stable circular orbits do exist at a radius $r_*=R_+$, where $R_+=\left(L^2+\sqrt{L^4-12L^2M^2}\right)/2M$, while unstable circular orbits exist at $r_*=R_{-}$, where $R_{-}=\left(L^2-\sqrt{L^4-12L^2M^2}\right)/2M$. A similar analysis of the motion of the particles in Newtonian mechanics in the presence of a central force field $f(r)$ was carried out in \cite{X2}. In nonrelativistic mechanics circular orbits in a central field do exist if the equation
\begin{equation}
V'(r)=-f(r)-\frac{L^2}{Mr^3}=0,
\end{equation}
 has real roots. Furthermore, if $r=r_0$ is a root of $V'(r)=0$, the circular orbit is stable if the condition
 \begin{equation}
 \left.V''(r)\right|_{r=r_0}=-\left.f'(r)\right|_{r=r_0}+\frac{3L^2}{Mr_0^4}>0,
 \end{equation}
is satisfied.

 In a realistic astrophysical environment, massive general relativistic objects, like, for example, black
holes or neutron stars, are often enclosed by an accretion disk. Accretion disks
around compact objects can be the basis of physical models that could convincingly provide explanations for many astrophysical phenomena, like, for example, active galactic nuclei and  X-ray binaries. The disks can be described theoretically  by assuming that they are composed of massive test particles (baryons) that evolve in the gravitational field of the central massive and compact astrophysical object. The disks cool down through the electromagnetic radiation emission from their surface, and this form of energy emission  represents an efficient physical mechanism for avoiding the extreme heating of the disk surface \cite{D1}. The disk has an inner edge, which is located at the marginally stable orbits of the gravitational, potential created by the central massive compact object. Hence, in higher orbits, the motion of the gas in the disk is Keplerian.

Therefore, the problem of the determination of the position of circular orbits, and of their stability, is fundamental from an astrophysical point of view. The electromagnetic emissivity properties of the accretion disks provide distinct observational signatures for different classes of astrophysical objects, including black holes, neutron, quark or other types of exotic stars. The parameters $l$ and $q$ of the GMGHS black hole can also be constrained from the physical properties of the accretion disks. The condition of the stability of the particle trajectories in the disk also imposes strong constraints on the parameters of central object. In this respect, the results obtained via the applications of the concept of Jacobi and Lyapunov stability may prove to be essential for the understanding of the nature of the black holes, or other types of compact objects.

For example, in \cite{D2}, it was shown that the equation governing the vertical perturbations of the trajectories of the test particles in the equatorial orbits around massive general relativistic objects is given by
\begin{equation}\label{lang}
\frac{d^{2}\delta z}{ds^{2}}+\nu \frac{d\delta z}{ds}+\omega _{\perp
}^{2}\delta z=\xi ^{z}\left[ g;z\right] ,
\end{equation}
where $\delta z$ is the perturbation of the coordinate $z$, $\nu$ is a constant, $\xi ^{z}\left[ g;z\right]$ is the external force, and
\begin{equation}
\omega _{\perp }^{2}=\left[ \Gamma _{tt,z}^{z}+2\Gamma _{t\phi
,z}^{z}\frac{\Omega }{c}+\Gamma _{\phi \phi
,z}^{z}\left(\frac{\Omega }{c}\right)^{2}\right] \left( u^{t}\right) ^{2},
\end{equation}
respectively. In the above equation $\Gamma _{\mu \nu}^{\lambda}$ denote the Christoffel symbols of a Riemannian metric, $\omega$ is the azimuthal angular velocity, while $u^t$ is the temporal component of the four-velocity of the particles in the disk.
To obtain Eq.~(\ref{lang}) it was assumed that the particles in the disk move along the geodesic lines. The presence of a viscous dissipation and of an external  (stochastic force) was also taken into account. Since the vertical perturbations of the disk are described by a second order differential equations, the study of the stability of the disk around the GMGHS black holes can be analyzed by using the theoretical concepts discussed in the present work.

To conclude, in the present study we have carried out  an
independent investigation of the stability of the geodesic trajectories of
massive, baryonic test particles moving in GMGHS geometry, by using both the Lyapunov
and the Jacobi methods for the stability analysis. We have obtained the basic mathematical the result that the
condition for Jacobi stability of circular orbits in a GMGHS spacetime is
the same as the condition for Lyapunov stability, meaning that in this case
these two types of stability are equivalent. This result is also a
consequence of the two-dimensional nature of the system of differential equations,
corresponding to the geodesic motion in the GMGHS geometry in spherical
static symmetry. For higher dimensional dynamical systems, and in the
presence of a complex behavior, the predictions of the Jacobi and Lyapunov
stability theories may be different, thus allowing for a better
explanation of the physical and mathematical properties of these systems on both qualitative and quantitative levels.

\vspace{6pt}




\section*{Acknowledgments}
We would like to thank the three anonymous reviewers for comments and suggestions that helped us to improve our manuscript.


\begin{thebibliography}{999}
\bibitem[Bofetta et al.(2002)]{1} G. Boffetta, M. Cencini, M. Falcioni, and A. Vulpiani. Predictability: a way to characterize complexity. {\em Physics Reports} {\bf 2002}, {\em 356}, 367--474.

\bibitem[Mancho et al.(2006)]{2}
A. M. Mancho, D. Small, and S. Wiggins.  A tutorial on dynamical systems concepts applied to Lagrangian transport in oceanic flows defined as finite time data sets: Theoretical and computational issues. {\em Physics Reports} {\bf 2006}, {\em 437}, 55--124.

\bibitem[Motter et al.(2013)]{3} A. E. Motter, M. Gruiz, G. K\'{a}rolyi, and T. T\'{e}l. Doubly Transient Chaos: Generic Form of Chaos in Autonomous Dissipative Systems. {\em Phys.Rev. Lett.} {\bf 2013}, {\em 111}, 194101.

\bibitem[Donetti et al. (2005)]{12} L. Donetti, P. I. Hurtado, and M. A. Munoz. Entangled Networks, Synchronization, and Optimal Network Topology. {\em
 Phys. Rev. Lett.} {\bf 2005}, {\em 95}, 188701.

\bibitem[Kosambi(1933)]{Ko33} D. D. Kosambi. Parallelism and path-spaces. {\em Mathematische Zeitschrift} {\bf 1933}, {\em 608}, 608--618.

\bibitem[Cartan(1933)]{Ca33} E. Cartan. Observations sur le m\'{e}emoir pr\'{e}c\'{e}dent. {\em Mathematische Zeitschrift} {\bf 1933}, {\em 37}, 619--622.

\bibitem[Chern(1939)]{Ch39} S. S. Chern. Sur la g\'{e}om\'{e}trie d’un syst\'{e}me d’equations differentialles du second ordre. {\em Bulletin des Sciences Math\'{e}matiques} {\bf 1939}, {\em 63}, 206--212.

\bibitem[Boehmer et al.(2012)]{rev} C. G. Boehmer, T. Harko, and S. V. Sabau. Jacobi stability analysis of dynamical systems -- applications in gravitation and cosmology.
 {\em Adv. Theor. Math.
Phys.} {\bf 2012}, {\em 16}, 1145--1196.

\bibitem[Antonelli(2003)]{An00} P. L. Antonelli (Editor), \textit{Handbook of Finsler geometry}, vol.
1, Kluwer Academic, Dordrecht, Holland, 2003.

\bibitem[Sabau(2005)]{Sa05a} S. V. Sabau. Systems biology and deviation curvature tensor. {\em Nonlinear Analysis: Real World Applications} {\bf 2005},
{\em 6}, 563--587.

\bibitem[Sabau(2005)]{Sa05} S. V. Sabau.  Some remarks on Jacobi stability.  {\em Nonlinear Analysis} {\bf 2005}, {\em 63}, 143--153.

\bibitem[Antonelli and Bucataru(2001)]{An93} P. L. Antonelli and I. Bucataru. New results about the geometric invariants in KCC theory, An. St. Univ. ”Al.I. Cuza” Iasi. Mat. (N.S.) {\bf 2001}, {\em 47}, 405--420.

\bibitem[Yajima and Nagahama(2007)]{YaNa07} T. Yajima and H. Nagahama. KCC-theory and geometry of the Rikitake system. {\em J. Phys. A: Math. Theor.} {\bf 2007}, {\em
40}, 2755--2772.

\bibitem[Harko and Sabau(2008)]{Har1} T. Harko and V. S. Sabau. Jacobi stability of the vacuum in the static spherically symmetric brane world models.
 {\em Phys. Rev. D} {\bf 2008},  {\em 77}, 104009.

\bibitem[Boehmer and Harko(2010)]{Har2} C. G. Boehmer and T. Harko. Nonlinear Stability Analysis of the Emden-Fowler Equation. {\em Journal of Nonlinear Mathematical
Physics} {\bf 2010}, {\em 17}, 503--516.

\bibitem[Yajima and Nagahama(2008)]{T0} T. Yajima and H. Nagahama. Nonlinear dynamical systems and KCC-theory. {\em Acta Mathematica Academiae
Paedagogicae Ny\'{\i}regyh\'{a}ziensis} {\bf 2008}, {\em 24}, 179--189.

\bibitem[Yajima and Nagahama(2010)]{T1} T. Yajima and H. Nagahama. Tangent bundle viewpoint of the Lorenz system and its chaotic behavior. {\em Physics Letters A} {\bf 2010}, {\em  374}, 1315--1319.

\bibitem[Abolghasem(2012)]{X1} H. Abolghasem. Liapunov stability versus Jacobi stability. {\em Journal of Dynamical Systems and Geometric
Theories} {\bf 2012}, {\em 10}, 13--32.

\bibitem[Abolghasem(2012)]{X2} H. Abolghasem. Jacobi Stability of Circular Orbits in a Central Force. {\em Journal of Dynamical Systems and Geometric
Theories} {\bf 2012}, {\em 10}, 197--214.

\bibitem[Abolghasem(2013)]{X3} H. Abolghasem.  Stability of circular orbits in Schwarzschild spacetime. {\em International Journal of Differential Equations
and Applications} {\bf 2013}, {\em 12}, 131--147.

\bibitem[Abolghasem(2013)]{X4} H. Abolghasem.  Jacobi stability of Hamiltonian systems. {\em International Journal of Pure and Applied Mathematics} {\bf 2013}, {\em 87}, 181--194.

\bibitem[Harko et al.(2015)]{Ha3} T. Harko, C. Y. Ho, C. S. Leung, and S. Yip. Jacobi stability analysis of the Lorenz system.
 {\em Int. J. of Geometric Methods in Modern Physics} {\bf 2015}, {\em 12}, 1550081.

\bibitem[Harko et al.(2015)]{Ha4} T. Harko, P. Pantaragphong, and S. Sabau. A new perspective on the Kosambi-Cartan-Chern theory, and its applications. {\em  arXiv:1509.00168} {\bf 2015}.

\bibitem[Harko et al.(2016)]{Ha4a}  T. Harko, P. Pantaragphong, and S. Sabau, Kosambi-Cartan-Chern (KCC) theory for higher-order dynamical systems. {\em International Journal of Geometric Methods in Modern Physics} {\bf 2016}, {\em 13}, 1650014.

\bibitem[Danila et al.(2016)]{Ha5} B. Danila, T. Harko, M. K. Mak, P. Pantaragphong, and S.
Sabau. Jacobi stability analysis of scalar field models with minimal coupling to gravity in a cosmological background. {\em  Advances in High Energy Physics} {\bf 2016}, {\em 2016}, 7521464.

\bibitem[Lake and Harko(2016)]{Ha6} M. J. Lake and T. Harko. Dynamical behavior and Jacobi stability analysis of wound strings. {\em The European Physical Journal C} {\bf 2016}
{\em 76}, 311.

\bibitem[Blaga et al.(2021)]{BBH} C. Blaga, P. A. Blaga, and T. Harko. Jacobi stability analysis of the classical restricted three body problem. {\em Romanian Astron. J. } {\bf 2021}, {\em 31}, 101--112.

\bibitem[Gibbons and Maeda(1988)]{GM88} G. W. Gibbons and K. Maeda. Black holes and membranes in higher-dimensional theories with dilaton fields.  {\em Nucl. Phys. B} {\bf 1988}, {\em 298}, 741--775.

\bibitem[Garfinkle et al.(1991)]{GHS91} T. Garfinkle, G.A. Horowitz and A. Strominger. Charged black holes in string theory. {\em Phys. Rev. D} {\bf 1991},
{\em 43}, 3140--3143.

\bibitem[Wainwright and Ellis(1997)]{cosm1} J. Wainwright and G. F. R. Ellis. {\em Dynamical Systems in
Cosmology}, Cambridge University Press, Cambridge, UK, 1997

\bibitem[Boehmer et al.(2012)]{cosm2} C. G. Boehmer, N. Chan, and R. Lazkoz. Dynamics of dark energy models and centre manifolds. {\em Phys. Lett. B} {\bf 2012},
{\em 714}, 11--17.

\bibitem[Boehmer and Chan(2014)]{cosm3} C. G. Boehmer and N. Chan. Dynamical systems in cosmology.  {\em arXiv:1409.5585} {\bf 2014}.

\bibitem[Garcia-Salcedo et al.(2015)]{cosm4} R. Garcia-Salcedo, T. Gonzalez, F. A. Horta-Rangel, I.
Quiros, and D. Sanchez-Guzm\'{a}n. Introduction to the application of dynamical systems theory in the study of the dynamics of cosmological models of dark energy. {\em Eur. J. Phys.} {\bf 2015}, {\em 36}, 025008.

\bibitem[Murray(1993)]{M} J. D. Murray. {\em  Mathematical Biology}, Springer Verlag, New York, 1993

\bibitem[Miron et al.(2001)]{MHSS} R. Miron, D. Hrimiuc, H. Shimada and V. S. Sabau. \textit{The
Geometry of Hamilton and Lagrange Spaces}, Kluwer Acad. Publ., Dordrecht;
Boston, 2001

\bibitem[Punzi and Wohlfarth(2009)]{Punzi} R. Punzi and M. N. R. Wohlfarth. Geometry and stability of dynamical systems.
 {\em Phys. Rev. E} {\bf 2009}, {\em 79}, 046606.

\bibitem[Blaga and Blaga(1998)]{BB98} C. Blaga and P. A. Blaga. On the geodesics for a spherically symmetric dilaton black hole. {\em  Serb. Astron. J.} {\bf 1998},  {\em 158}, 55--59.

\bibitem[Fernando (2012)]{F12} S. Fernando. Null geodesics of charged black holes in string theory. {\em Phys. Rev. D} {\bf 2012}, {\em 85}, 024033.

\bibitem[Pradhan(2015)]{P12} P.P. Pradhan. Horizon straddling ISCOs in spherically symmetric string black holes. {\em Int. J. Mod. Phys. D} {\bf 2015}, {\em 24}, 1550086.

\bibitem[Olivares and Villanueva(2013)]{ov13} M. Olivares and J. R. Villanueva. Massive neutral particles on heterotic string theory. {\em Eur. Phys. J. C} {\bf 2013}, {\em 73},  2659.

\bibitem[Blaga(2013)]{B13} C. Blaga. Circular time-like geodesics around a charged spherically symmetric dilaton black hole. {\em Automat. Comp. Appl. Math.} {\bf 2013}, {\em 22}, 41--49.

\bibitem[Blaga(2015)]{B15} C. Blaga. Timelike geodesics around a charged spherically
symmetric dilaton black hole. {\em Serb. Astron. J.} {\bf 2015}, {\em 190}, 41--48.

\bibitem[Shahidi et a.(2020)]{D1} S. Shahidi, T. Harko, and Z. Kov\'{a}cs. Distinguishing Brans-Dicke-Kerr type naked singularities and black holes with their thin disk electromagnetic radiation properties. {\em Eur. Phys. J. C} {\bf 2020}, {\em 80}, 162.


\bibitem[Harko and Mocanu(2012)]{D2} T. Harko and G. R. Mocanu. Stochastic oscillations of general relativistic discs. {\em MNRAS} {\bf 2012}, {\em 421}, 3102--3110.


\end{thebibliography}
\end{document}